\documentclass[aps,prl,twocolumn,groupedaddress,notitlepage,%showpacs,
floatfix,superscriptaddress]{revtex4-2}

\usepackage{graphicx,graphics,epsfig,subfigure,times,bm,bbm,amssymb,amsmath,amsthm,mathrsfs,MnSymbol} \usepackage{gensymb} \usepackage{amsfonts} \usepackage{float} \usepackage[matrix,frame,arrow]{xypic} \usepackage[pdfstartview=FitH]{hyperref} %\usepackage{subfigure}
\usepackage{times}
\usepackage{float}
\usepackage{graphics}
\usepackage[T1]{fontenc}

\usepackage{braket}  %Dirac Notation in QM
\usepackage{enumerate} \usepackage[normalem]{ulem}
\hypersetup{ colorlinks=true,       % false: boxed links; true: colored links
	linkcolor=red,          % color of internal links
	citecolor=blue,        % color of links to bibliography
	filecolor=magenta,      % color of file links
	urlcolor=blue,           % color of external links
	runcolor=cyan }

\begin{document}
	\title{Probing Operator Spreading via Floquet Engineering in a Superconducting Circuit}
	
	\author{S. K.~Zhao}
	\altaffiliation[]{These authors contributed equally to this work.}
	\affiliation{Beijing National Laboratory for Condensed Matter Physics, Institute of
		Physics, Chinese Academy of Sciences, Beijing 100190, China} \affiliation{School of
		Physical Sciences, University of Chinese Academy of Sciences, Beijing 100190, China}
	\affiliation{Beijing Academy of Quantum Information Sciences, Beijing 100193, China}
	
	\author{Zi-Yong~Ge}
	\altaffiliation[]{These authors contributed equally to this work.}
	\affiliation{Beijing National Laboratory for Condensed Matter Physics, Institute of
		Physics, Chinese Academy of Sciences, Beijing 100190, China} \affiliation{School of
		Physical Sciences, University of Chinese Academy of Sciences, Beijing 100190, China}
	
	\author{Zhongcheng~Xiang}
	\altaffiliation[]{These authors contributed equally to this work.}
	\affiliation{Beijing National Laboratory for Condensed Matter Physics, Institute of
		Physics, Chinese Academy of Sciences, Beijing 100190, China}
	
	\author{G. M.~Xue} \affiliation{Beijing Academy of Quantum Information Sciences, Beijing
		100193, China}
	
	\author{H. S.~Yan} \affiliation{Beijing National Laboratory for Condensed Matter Physics,
		Institute of Physics, Chinese Academy of Sciences, Beijing 100190, China}
	\affiliation{School of Physical Sciences, University of Chinese Academy of Sciences,
		Beijing 100190, China}
	
	\author{Z. T.~Wang} \affiliation{Beijing National Laboratory for Condensed Matter
		Physics, Institute of Physics, Chinese Academy of Sciences, Beijing 100190, China}
	\affiliation{School of Physical Sciences, University of Chinese Academy of Sciences,
		Beijing 100190, China}
	
	\author{Zhan~Wang} \affiliation{Beijing National Laboratory for Condensed Matter Physics,
		Institute of Physics, Chinese Academy of Sciences, Beijing 100190, China}
	\affiliation{School of Physical Sciences, University of Chinese Academy of Sciences,
		Beijing 100190, China}
	
	\author{H. K.~Xu} \affiliation{Beijing Academy of Quantum Information Sciences, Beijing
		100193, China}
	
	\author{F. F.~Su} \affiliation{Beijing National Laboratory for Condensed Matter Physics,
		Institute of Physics, Chinese Academy of Sciences, Beijing 100190, China}
	
	\author{Z. H.~Yang} \affiliation{Beijing National Laboratory for Condensed Matter
		Physics, Institute of Physics, Chinese Academy of Sciences, Beijing 100190, China}
	\affiliation{School of Physical Sciences, University of Chinese Academy of Sciences,
		Beijing 100190, China}
	
	\author{He~Zhang} \affiliation{Beijing National Laboratory for Condensed Matter Physics,
		Institute of Physics, Chinese Academy of Sciences, Beijing 100190, China}
	\affiliation{School of Physical Sciences, University of Chinese Academy of Sciences,
		Beijing 100190, China}
	
	\author{Yu-Ran~Zhang} \affiliation{Theoretical Quantum Physics Laboratory, RIKEN Cluster
		for Pioneering Research, Wako-shi, Saitama 351-0198, Japan}
	
	\author{Xue-Yi~Guo} \affiliation{Beijing National Laboratory for Condensed Matter
		Physics, Institute of Physics, Chinese Academy of Sciences, Beijing 100190, China}
	
	\author{Kai~Xu} \affiliation{Beijing National Laboratory for Condensed Matter Physics,
		Institute of Physics, Chinese Academy of Sciences, Beijing 100190, China}
	\affiliation{Beijing Academy of Quantum Information Sciences, Beijing 100193, China}
	\affiliation{CAS Center for Excellence in Topological Quantum Computation, UCAS, Beijing
		100190, China}
	
	\author{Ye~Tian} \affiliation{Beijing National Laboratory for Condensed Matter Physics,
		Institute of Physics, Chinese Academy of Sciences, Beijing 100190, China}
	
	\author{H. F.~Yu} \email{hfyu@baqis.ac.cn} 
	\affiliation{Beijing Academy of Quantum Information Sciences, Beijing 100193, China}
	
	\author{D. N.~Zheng} \email{dzheng@iphy.ac.cn} \affiliation{Beijing National Laboratory
		for Condensed Matter Physics, Institute of Physics, Chinese Academy of Sciences, Beijing
		100190, China} \affiliation{School of Physical Sciences, University of Chinese Academy of
		Sciences, Beijing 100190, China} \affiliation{CAS Center for Excellence in Topological
		Quantum Computation, UCAS, Beijing 100190, China} \affiliation{Songshan Lake Materials
		Laboratory, Dongguan 523808, China}
	
	\author{Heng~Fan} \email{hfan@iphy.ac.cn} \affiliation{Beijing National
	Laboratory for Condensed Matter Physics, Institute of Physics, Chinese Academy
	of Sciences, Beijing 100190, China} \affiliation{School of Physical Sciences,
	University of Chinese Academy of Sciences, Beijing 100190, China}
\affiliation{Beijing Academy of Quantum Information Sciences, Beijing 100193,
	China} \affiliation{CAS Center for Excellence in Topological Quantum
	Computation, UCAS, Beijing 100190, China} \affiliation{Songshan Lake Materials
	Laboratory, Dongguan 523808, China}
		
	\author{S. P.~Zhao} \email{spzhao@iphy.ac.cn} \affiliation{Beijing National Laboratory
		for Condensed Matter Physics, Institute of Physics, Chinese Academy of Sciences, Beijing
		100190, China} \affiliation{School of Physical Sciences, University of Chinese Academy of
		Sciences, Beijing 100190, China} \affiliation{CAS Center for Excellence in Topological
		Quantum Computation, UCAS, Beijing 100190, China} \affiliation{Songshan Lake Materials
		Laboratory, Dongguan 523808, China}

\begin{abstract} 
Operator spreading, often characterized by out-of-time-order correlators
(OTOCs), is one of the central concepts in quantum many-body physics. However,
measuring OTOCs is experimentally challenging due to the requirement of
reversing the time evolution of systems. Here we apply Floquet engineering to
investigate operator spreading in a superconducting 10-qubit chain. Floquet
engineering provides an effective way to tune the coupling strength between
nearby qubits, which is used to demonstrate quantum walks with tunable couplings,
reversed time evolution, and the measurement of OTOCs. A clear light-cone-like
operator propagation  is observed in the system with multiple excitations, and
has a nearly equal velocity as the single-particle quantum walk. For the
butterfly operator that is nonlocal (local) under the Jordan-Wigner
transformation, the OTOCs show distinct behaviors with (without) a signature  of
information scrambling in the near integrable system.
\end{abstract}
	
\maketitle

\textit{Introduction.}---%
Spreading of quantum operators,  probed by out-of-time-order correlators
(OTOCs)~\cite{Shenker2014,Roberts2015,Hosur2016,Shen2017,
	Swingle2018,Lin2018a,Lin2018b,Joshi2020}, is a new viewpoint to study the
dynamics of quantum many-body systems. For instance, it is closely related to
the Lieb-Robinson bound~\cite{Lieb1972,Bravyi2006} in local quantum systems and
information scrambling in quantum chaotic systems
~\cite{Shenker2014,Roberts2015,Hosur2016,Kitaev2015,Bohrdt2017,Nahum2018,Keyserlingk2018}. 
Given two local operators $\hat W$ and $ \hat V$, the OTOC with a pure state $\ket{\psi_0}$ 
is defined as~\cite{Swingle2018}
\begin{align} 
\label{otoc} C(t) = \bra{\psi_0}\hat {W}(t) \hat{ V }\hat{ W}^\dagger (t) \hat{ V}^\dagger\ket{\psi_0}, 
\end{align} 
where $\hat{ W}(t) = e^{i\hat{ H} t}\hat{ W} e^{-i\hat{ H} t}$ is the butterfly
operator, and $\hat H$ is the system Hamiltonian. The OTOC is related
	to the commutator between $\hat {W}(t)$ and $\hat{V}$. In a
	dynamical process, it usually starts from unity where two operators fully
	commute, and decays when they no longer commute and the system may become
	scrambled~\cite{Lin2018a,Lin2018b}. To probe $C(t)$, one generally needs to
reverse the time evolution of the
system~\cite{Shen2017,Li2017,Nie2020,Garttner2017,Landsman2019,Mi2021,Braum2021}, 
which is experimentally very challenging~\cite{rem}. In the \emph{digital}
quantum simulation paradigm~\cite{Buluta2009,Georgescu2014}, the dynamics of
quantum many-body systems can be digitalized via a series of quantum gates
according to the Trotter-Suzuki decomposition, and the reversed time evolution
can be easily implemented in a similar way. These have been
	investigated in various physical systems such as nuclear magnetic resonance
	~\cite{Li2017,Nie2020}, trapped ions~\cite{Garttner2017,Landsman2019}, and
	superconducting circuits~\cite{Mi2021}. On the other hand, it is efficient to
encode the reversed  dynamics of the target system with a fully-controllable
quantum simulator, which is a task of \emph{analog} quantum
simulations~\cite{Buluta2009,Georgescu2014}. In the system with special
symmetry, the reversal of time evolution is realized with a combined
\emph{digital-analog} scheme in superconducting circuits~\cite{Braum2021}, which
nevertheless cannot be simply generalized to other systems without the
corresponding symmetry. Hence a method to encode the reversed time evolution of
a quantum many-body system in a \emph{full analog} quantum simulation paradigm
is highly desirable.

Floquet engineering, using time-periodic driving, is  a powerful tool for the
coherent manipulation of quantum many-body states and the control of their
dynamics. It has achieved a great success in cold atom
systems~\cite{Eckardt2017} for a number of studies~\cite{Dunlap1986,Lignier2007,Eckardt2009,Dalibard2011,Jotzu2014}. 
It has also been applied in superconducting circuits for realizing qubit
switch~\cite{Wu2018}, high-fidelity quantum gates~\cite{Reagor2018}, quantum
state transfer~\cite{Li2018}, and the model of topological magnon
insulators~\cite{Cai2019}. Floquet driving can be used to tune both the
magnitude and phase of the coupling  between nearby qubits, thus offering a
possible way for reversing the dynamics of a quantum system~\cite{Eckardt2017}, 
measuring OTOCs~\cite{Shen2017}, and probing operator spreading.

\begin{figure*}[t] \includegraphics[width=0.9\textwidth]{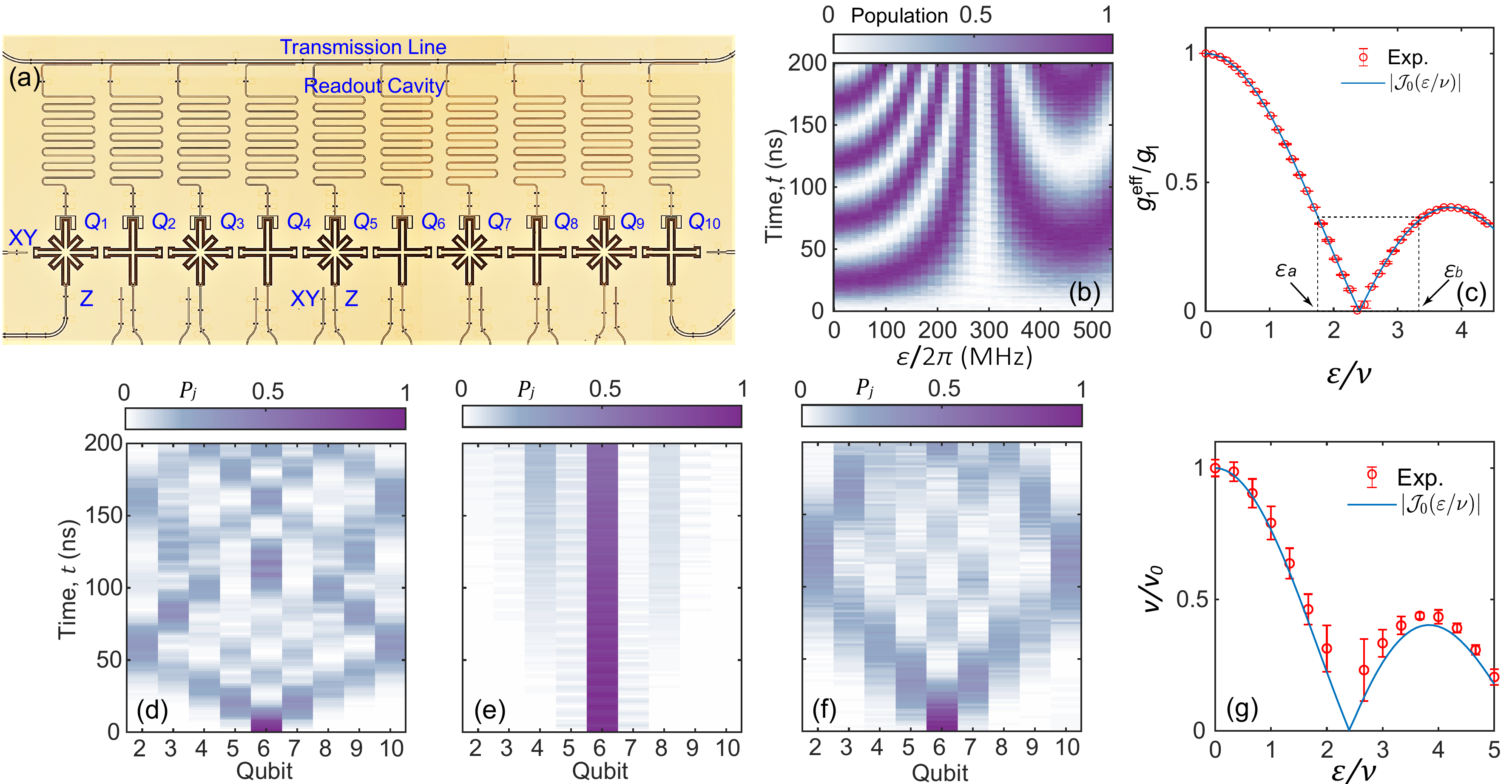} 
	\caption{Device and quantum walk with tunable coupling. (a) Optical micrograph
		of the  superconducting processor containing 10 transmon qubits arranged into a
		chain. Each qubit has a microwave line for the $XY$ control, a flux bias line
		for the $Z$ control, and a readout resonator for measurement. The NN qubits are     
		coupled capacitively with almost the same strength~\cite{SM}. The parameters of
		the device are presented in the Supplemental Material (SM) in detail~\cite{SM}.
		(b) Experimental results of excitation oscillation between $Q_1$ and $Q_2$.
		The two qubits are prepared in the initial state of $\ket{01},$ and ac
		magnetic flux is applied on $Q_1$ with an amplitude $\varepsilon$. The color
		bar represents the expectation value of the photon number of $Q_1$. (c)
		Experimental coupling strength $g_{1}^{\text{eff}}$ versus ac magnetic flux
		amplitude $\varepsilon$, which satisfies $g_{j}^{\text{eff}}/g_1 \approx
		\mathcal{J}_0(\varepsilon/\nu)$. Two arrows indicate the values,
		$\varepsilon_a$ and $\varepsilon_b$, used in the measurement of OTOCs for the
		forward and backward time evolution, respectively.  Single-photon quantum walks
		for (d) $\varepsilon/2\pi=120$~MHz, (e) $\varepsilon/2\pi=288.6$~MHz, and (f)
		$\varepsilon/2\pi=400$~MHz, observed on $Q_2$-$Q_{10}$ with the initial state
		$\ket{000010000}$.  (g) Normalized group velocity $v/v_0$ of photon propagation
		versus $\varepsilon/\nu$, where $v_0= 117\pm4$ sites/$\mu$s corresponds to the
		value with $\varepsilon=0$~\cite{SM}. The error bar is estimated by fitting
		errors. Each point is the average result of $8,000$ single-shot readouts.}
	\label{fig_1}
\end{figure*}

In this Letter, we present a systematic study of operator spreading using
Floquet engineering in a 1D array of 10 superconducting qubits. By precisely
adjusting the ac magnetic flux at specific qubits, we are able to tune the
coupling strength of nearby qubits and demonstrate the
single-photon quantum walk with varying coupling, reversed time evolution, and
the measurement of OTOCs. The Pauli operators $\hat \sigma^z$ and
	$\hat \sigma^x$ are taken as butterfly operators, which are local and
	nonlocal under the Jordan-Wigner transformation, respectively. We show that, 
	with multiple excitations, a clear light-cone-like operator
propagation can be observed with a velocity nearly equal to the group velocity
of single-photon quantum walks. The OTOC with $\hat{W}$ = $\hat
	\sigma^z$ shows a revival to the initial value after a quick decay and finally
	approaches zero, which  characterizes the nonthermalized process in the
	absence of scrambling. In contrast, no such revival is observed for the OTOC
	with $\hat{W}$ = $\hat \sigma^x$, showing a signature of scrambling. Our 
	results demonstrate distinct behaviors of OTOCs in the nearly integrable system,
	some of which resemble those in nonintegrable chaotic
	systems~\cite{Lin2018a,Lin2018b}.

\textit{Experimental setup and protocol.}---%
Superconducting qubits can be manipulated individually as well as simultaneously,
which provide a convenient platform for simulating quantum many-body
systems~\cite{Xu2018,Roushan2017,Salathe2015,Barends2015,Zhong2016,Flurin2017,Song2018, Ma2019,Yan2019,Ye2019,Guo2019,Arute2019,Xu2020,Guo2021,Guoqj2021}. 
Our experiment is performed on a 1D array of 10  coupled transmon qubits, shown in Fig.~\ref{fig_1}(a). In the rotating frame with a common  frequency, 
the system is described by the 1D Bose-Hubbard model~\cite{Roushan2017,Yan2019,Ye2019}
\begin{align}
	\label{Hxy} \hat{H}(t) =&\sum_{j=1}^{9}g_{j}(\hat{a}^{\dagger}_{j}
	\hat{a}_{j+1}+\text{h.c.})+ \sum_{j=1}^{10} \omega_{ j}(t)\hat{n}_{j} ~~\nonumber\\
	&+\sum_{j=1}^{10}\frac{U_j}{2} \hat{n}_{j}(\hat{n}_{j}-1), 
\end{align} 
where $\hat{a}^{\dagger}_{j}$ ($\hat{a}_{j}$) is the bosonic creation
(annihilation) operator, $\hat{n}_{j}$ = $\hat{a}^{\dagger}_{j}\hat{a}_{j}$ is
the number operator, $U_j$ is the on-site interaction, and $g_{j}$ is the
nearest-neighbor (NN) coupling strength. In our experiment, we bias the qubit
frequency with ac magnetic flux, i.e., we have $\omega_{ j}(t) =\varepsilon_j
\cos(\nu t)$, with $\nu$ and $\varepsilon_j$ being the ac frequency and
amplitude, respectively. Thus $\hat H(t)$ describes a Floquet
system satisfying $\hat{H}(t)=\hat{H}(t+T)$ with a period of $T=2\pi/\nu$
~\cite{SM}. When $\nu \gg g_j$, under the hard-core boson approximation with
$|U|\gg g_j$, we obtain an effective time-independent
Hamiltonian~\cite{Eckardt2017}
\begin{align}
	\label{Heff} \hat{H}_{\text{eff}} =&\sum_{j=1}^{9}g_{j}^{\text{eff}}(\hat{\sigma}^+_{j}
	\hat{\sigma}^-_{j+1}+\hat{\sigma}^+_{j+1} \hat{\sigma}^-_{j})~, 
\end{align} 
where $\hat{\sigma}^{\pm} = (\hat{\sigma}^x\pm i\hat{\sigma}^y)/2$ with $\hat{\sigma}^{x,y,z}$
being Pauli matrices. The effective coupling strength has the form~\cite{Eckardt2017}
\begin{align} 
	\label{Jeff} g_{j}^{\text{eff}} \approx
	g_j\mathcal{J}_0\big(\frac{\varepsilon_j-\varepsilon_{j+1}}{\nu}\big), 
\end{align} 
where $\mathcal{J}_0\big(x\big)$ is the Bessel function of order zero. 

We tune the effective coupling strength between NN
qubits by changing $\varepsilon_j$ with fixing $\nu/2\pi$ = 120~MHz.
Figure~\ref{fig_1}(b) shows the experimental results of the excitation
oscillation between $Q_1$ and $Q_2$, where the $\varepsilon$-dependent
oscillation period can be seen. Using the Fourier transformation, we obtain the
effective coupling strength as a function of $\varepsilon$, plotted in
Fig.~\ref{fig_1}(c), which fits well with
Eq.~(\ref{Jeff}). In order to have a common coupling strength between each NN
qubit pair, we only drive the odd qubits
 with the same amplitude $|\varepsilon_j|=\varepsilon$, so the coupling
strength approximates $g_{j}\mathcal{J}_0(\varepsilon/\nu)$. In addition, we
stagger the phase of the applied flux with $\varepsilon_1,
\varepsilon_5,\varepsilon_9=\varepsilon$ and
$\varepsilon_3,\varepsilon_7=-\varepsilon$ to partly reduce the unwanted
next-nearest-neighbor (NNN) coupling~\cite{SM}. Hence we are able to set
identical coupling strength for each NN qubit pair with adjustable values from
positive to negative.

\textit{Quantum walks with tunable coupling.}---%
The quantum walk is a fundamental process for quantum simulation and
computation~\cite{Yan2019,Ye2019,Gong2021,Underwood2012,Childs2013}.
Figures~\ref{fig_1}(d)-(f) show the experimental observation in a 9-qubit system
($Q_2-Q_{10}$) with varying effective coupling strength. The experiment starts with all
qubits biased at their idle points, and the central qubit $Q_6$  is excited from
the ground state $\ket{0}$ to the first-excited state $\ket{1}$ by an $X_\pi$
gate to prepare an initial state $\ket{000010000}$~\cite{SM}. Afterwards, they
are biased to the same frequency and the periodic driving is applied. The system
then evolves with almost homogeneous coupling strength between NN qubits. The photon
density distribution $P_j$ is measured after the system's evolving for a time
$t$, where $P_j(t):=\bra{\psi(t)}\hat{\sigma}^+_{j}\hat{\sigma}^-_{j}\ket{\psi(t)}$
with $\ket{\psi(t)}$ being the wave function at time $t$. Figure~\ref{fig_1}(d)
shows the result for $\varepsilon/2\pi=120$~MHz, which displays a single-photon 
light-cone-like propagation~\cite{Yan2019,Ye2019} in both the left and right directions.

\begin{figure}[t] \includegraphics[width=0.45\textwidth]{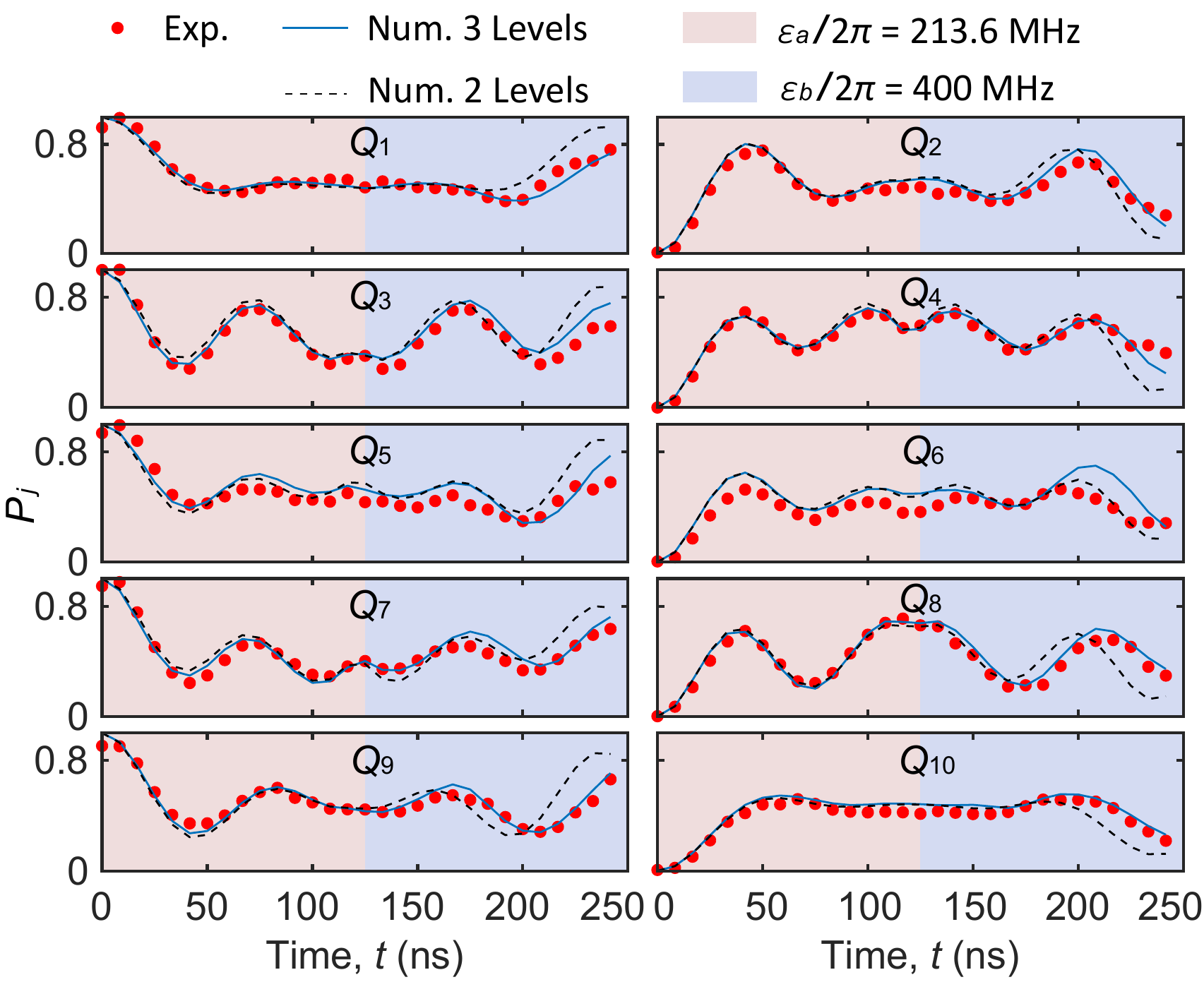} 
	\caption{Reversed time evolution of the 10-qubit chain with the initial state
		$\ket{1010101010}$, realized by using $\varepsilon_a$ and $\varepsilon_b$ for
		the time before and after 125~ns, respectively. Symbols are the experimental
		results. Solid and dashed lines are the calculated results based on the
		Hamiltonian in Eq.~(\ref{Hxy}) by considering three and two energy levels for each
		qubit, respectively.}
	\label{fig_2} 
\end{figure}

We find that the propagation velocity $v$ decreases with increasing
$\varepsilon$ and vanishes at about $\varepsilon/2\pi\approx288.6$~MHz ($\varepsilon/\nu\approx2.4$),
corresponding to the first zero of $\mathcal{J}_0$. At this point, the effective NN coupling reduces
to zero, leading to dynamic
localization~\cite{Dunlap1986,Lignier2007,Eckardt2009} as illustrated in
Fig.~\ref{fig_1}(e). Further increasing $\varepsilon$ results in reappearing of
the linear propagation, see Fig.~\ref{fig_1}(f) for the case of
$\varepsilon/2\pi=400$~MHz.  In Fig.~\ref{fig_1}(g), we show the normalized
group velocity $v/v_0$ versus $\varepsilon/\nu$, where $v_0$ = 117$\pm4$
sites/$\mu$s is the value at $\varepsilon=0$. It can be seen that the data are also well described by the 0-th Bessel function, and the difference
mainly originates from the unwanted NNN coupling~\cite{SM}. Specifically, the group
velocity for $\varepsilon/2\pi=400$~MHz is $v=46\pm4$~sites/$\mu$s, which will
be compared with the operator spreading velocity below. These
results indicate a free single-photon propagation~\cite{Yan2019,Ye2019} that is
limited by the Lieb-Robinson bound~\cite{Lieb1972,Bravyi2006}.

\textit{Reversed time evolution.}---%
Reversing the dynamics of a quantum many-body system is of great interest and
facilitates the OTOC measurement~\cite{Swingle2018,Braum2021}. Equation~(\ref{Jeff})
indicates that the effective coupling $g_{j}^{\text{eff}}$ can be positive or
negative. For single-photon quantum walks, the last term in Eq.~(\ref{Hxy})
vanishes, so the system is governed by the Hamiltonian~(\ref{Heff}). In this case, the time evolution can be precisely
reversed by changing the sign of the Hamiltonian. However, 
OTOCs provide a technique for studying  operator spreading and quantum
information propagation in systems with multi-particle filling, which
 cannot be probed via single-particle propagation. Hence we will focus
on the multi-photon system and discuss the effect of
high-level occupations.

To observe the reversed time evolution, we choose the 
 initial state as the N\'{e}el state $\ket{1010101010}$. The system evolves with a driving
amplitude $\varepsilon=\varepsilon_a= 213.6$~MHz for the first 125~ns and with
$\varepsilon=\varepsilon_b= 400$~MHz for the last 125~ns. Here we have
$\mathcal{J}_0(\varepsilon_a/\nu) =-\mathcal{J}_0(\varepsilon_b/\nu)$, 
corresponding to $g_{j}^{\text{eff}} \approx\pm 4$~MHz for $\varepsilon_a$ and
$\varepsilon_b$, respectively, as indicated by arrows in Fig.~\ref{fig_1}(c).
The experimental results, plotted in Fig.~\ref{fig_2} as symbols,  are fairly reproduced by 
numerical simulations (lines). We can see that $P_j(t)$ is
nearly symmetric about $t=125$~ns during the time evolution, demonstrating  that two
effective time-independent Hamiltonians are almost opposite in sign.

\begin{figure}[t] \includegraphics[width=0.45\textwidth]{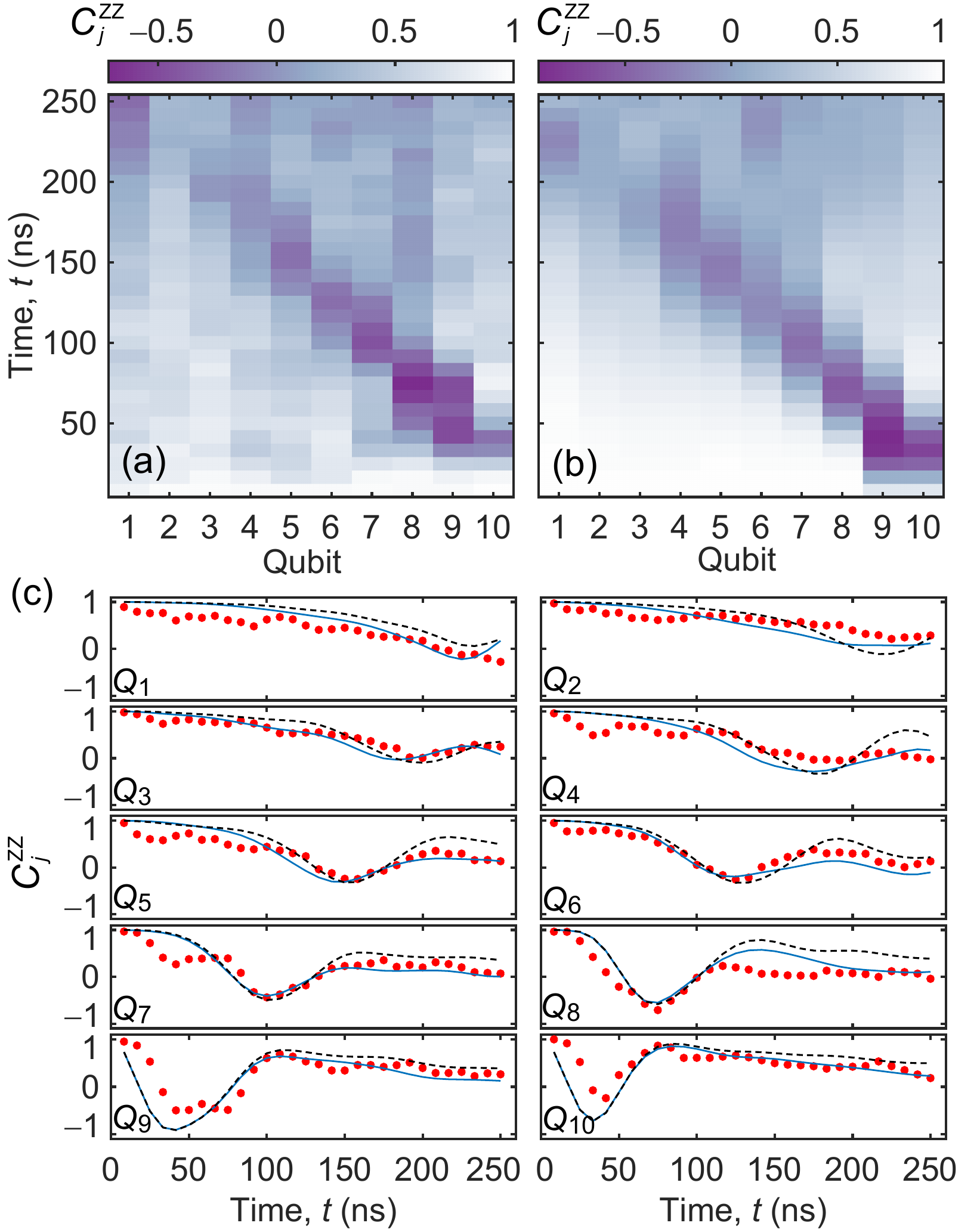} 
	\caption{(a) Experimental results of $C_j^{\text{ZZ}}$, with the initial state
	$\ket{0101010101}$ and operators $\hat W = \hat \sigma^z_{10}$, $\hat V = \hat
	\sigma^z_j$ ($j$ = 1, ..., 10). The post-selection is applied due
		to the $U(1)$ symmetry, i.e., only the single-shot results that
		have the same number of bosons as the initial state are considered. (b) Numerical results
	considering ZZ interaction between NN qubits. (c) Comparison between
	experimental $C_j^{\text{ZZ}}$ (symbols) and two calculated results using $\hat
	H_{\text{eff}}$ (dashed lines) and considering ZZ interaction (solid lines).}
\label{fig_3} 
\end{figure}

\textit{OTOCs and operator spreading.}---%
To measure OTOCs and probe operator spreading, we consider two cases
\cite{Lin2018a,Lin2018b}. First, we choose $\hat W = \hat
\sigma^z_{10}$, $\hat V = \hat \sigma^z_j$ ($j$ = 1,...,10), and the initial
state is $\ket{\psi_1}=\ket{0101010101}$. Since $\hat V \ket{\psi_1}=\hat
\sigma^z_j \ket{\psi_1}=-(-1)^j \ket{\psi_1}$, from Eq.~(\ref{otoc}), the
corresponding OTOC reads 
$C_j^{\text{ZZ}} (t) = -(-1)^j \bra{\phi_z(t)}\hat \sigma^z_j\ket{\phi_z(t)}$,
where $\ket{\phi_z(t)}=\hat \sigma^z_{10}(t)\ket{\psi_1}$~\cite{Nie2020}.  Second,
we let $\hat W = \hat \sigma^x_{10}$, $\hat V = \hat \sigma^x_j$ ($j$ =
1,...,10), and the initial state is $\ket{\psi_2}=\ket{++++++++++}$, where
$\ket{+}$ is the eigenstate of $\hat \sigma^x$ with an eigenvalue $+1$. Similarly, the
OTOC is measured as 
$C_j^{\text{XX}} (t) = \bra{\phi_x (t)}\hat \sigma^x_j
	\ket{\phi_x(t)}$,
where $\ket{\phi_x(t)}=\hat \sigma^x_{10}(t)\ket{\psi_2}$~\cite{Nie2020}. In the experiment,
we first set $\varepsilon=\varepsilon_a$ and let the system evolve for a time
$t$  from the initial state. Then we apply a $Z$ (or $X$) gate
on $Q_{10}$ and let the system evolve reversely  for time $t$ by setting
$\varepsilon=\varepsilon_b$. Finally, we measure the observable $\hat
\sigma^z_j$ (or $\hat \sigma^x_j$) to evaluate 
OTOC.

The experimental and numerical results of $C_j^{\text{ZZ}}$ are presented in
Fig.~\ref{fig_3}. We can observe a clear light-cone-like operator propagation in
Fig.~\ref{fig_3}(a). The velocity of operator spreading is calculated to be
$40\pm7$~sites$/\mu$s~\cite{SM}, which almost equals to the group velocity of
single-photon quantum walks with the identical coupling strength.
The small difference between these two velocities mainly comes from the
NNN coupling, which enlarges the group velocity of single-photon quantum
walks~\cite{SM}. These results demonstrate that operator spreading with the
local Hamiltonian is also limited by the Lieb-Robinson bound. The measured OTOCs
are also plotted in Fig.~\ref{fig_3}(c) as symbols. They show a clear decay at
the early stage and then revive almost back to the initial value of +1 for
qubits near $Q_{10}$, which implies the absence of information scrambling. This
can be explained considering that the effective Hamiltonian in Eq.~(\ref{Heff})
is integrable~\cite{Hosur2016,Li2017}. It maps to free fermions under the
Jordan-Wigner transformation and the $\hat \sigma^z$ operator does not change
under the map. Finally, $C_j^{\text{ZZ}}$ gradually decays to zero at later
time, indicating that $\ket{\phi_z(t)}$ tends to a steady state. Since the
initial state is in the half-filling sector and $\ket{\phi_z(t)}$ is spin
conserved, $C_j^{\text{ZZ}}$ represented by spin distribution will finally
stabilize at zero.

\begin{figure}[t] \includegraphics[width=0.45\textwidth]{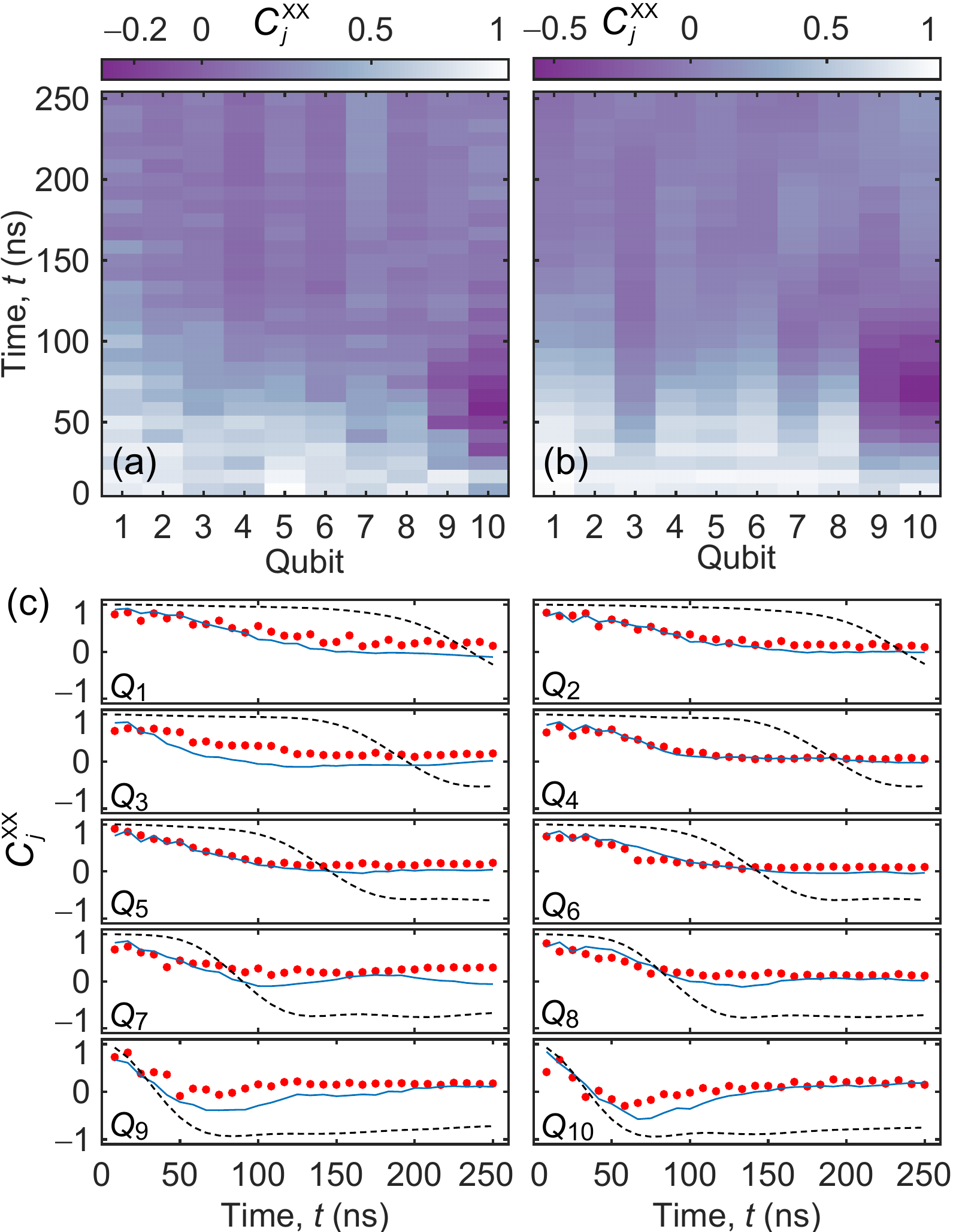} 
	\caption{(a) Experimental results of $C_j^{\text{XX}}$ with the initial state
	$\ket{++++++++++}$ and operators $\hat W = \hat \sigma^x_{10}$, $\hat V = \hat
	\sigma^x_j$ ($j$ = 1, ..., 10). (b) Numerical results
	considering the qubit third energy level. (c) Comparison between experimental
	$C_j^{\text{XX}}$ (symbols) and two numerical results using two (dashed
	lines) and three (solid lines) qubit levels.}
	\label{fig_4} 
\end{figure}

These results show that OTOCs can well characterize the nonthermalized process
and operator spreading in the multi-particle system. 
The effect of the last term in Eq.~(\ref{Hxy}) can also be seen, where it is not time reversible
under the Floquet driving leading to a discrepancy from  ideal time reversal~\cite{Braum2021}. For the results in Fig.~\ref{fig_2},
numerical simulations show that the population of the second-excited state
varies with a maximum value up to 10$\%$, as compared to those of the
first-excited state around 50$\%$. In Fig.~\ref{fig_2}, we present two results
calculated by three- and two-level approximation of $\hat H(t)$, respectively.
The result considering three levels is closer to experimental data. For the OTOC
results in Fig.~\ref{fig_3}, where the post-selection is used, we take into
account the influence of the third level through ZZ
interaction~\cite{McKay2017,Braum2021}.  In Fig.~\ref{fig_3}(c), we can find
that this numerical result is closer to experimental data than those calculated
using $\hat H_{\text{eff}}$ in Eq.~(\ref{Heff})~\cite{SM}.

The corresponding results of $C_j^{\text{XX}}$  are shown in Fig.~\ref{fig_4}. A
significant difference from $C_j^{\text{ZZ}}$ is that no similar revival back to
+1 is observed, which suggests the presence of scrambling with the $\hat
\sigma^x$ butterfly operator~\cite{Lin2018a,Lin2018b}. In the results, several
features are blurred due to the high level participation~\cite{Braum2021}. The
solid lines in Fig.~\ref{fig_4}(c) are the results calculated by considering the
qubit third level~\cite{SM}, showing a satisfactory fit to experimental data. In
addition, the results calculated by using two levels (dashed lines) demonstrate
clear properties that $C_j^{\text{XX}}$ would have: First, the wavefront of OTOC
can be better defined and propagates to the other edge of the qubit chain at a
time nearly equal to that for $C_j^{\text{ZZ}}$. Second, $C_j^{\text{XX}}$
decreases at early times from +1 to -1 and almost retains in the rest of
 time range up to 250 ns. Here, the difference between butterfly
operators $\hat \sigma^z$ and $\hat \sigma^x$ is that the former is local under
the Jordan-Wigner transformation, whereas the latter maps to nonlocal
Pauli-string operators giving rise to the behavior characteristic of
scrambling~\cite{Lin2018a,Lin2018b}. We emphasize that the scrambling here is
fundamentally different from that in nonintegrable systems, in which it persists
in a long timescale~\cite{Hosur2016}. For both integrable and nonintegrable systems, the length
of the string operators would increase linearly until approaching the system
size. Afterwards, it will start to decrease and saturate for the two systems,
respectively~\cite{Roberts2015}. In SM~\cite{SM}, we show the periodic
behavior of $C_j^{\text{XX}}$ by numerical simulation in a larger timescale,
which demonstrates the existence of string operator shrinking in our near
integrable system.

\textit{Summary and outlook.}---%
We have used Floquet engineering, a full analog method, for probing operator
spreading. Quantum walks with tunable coupling, reversed time evolution, and the
measurement of OTOCs were demonstrated in a superconducting qubit chain. 
We observed a linear propagation of quantum operator with a velocity
nearly equal to the group velocity of single-photon quantum walk, and also found
that the OTOCs behave differently between $\hat \sigma^z$ and $\hat
\sigma^x$ butterfly operators. The method may have further applications
for the simulation of quantum many-body physics, e.g., quantum information
scrambling and thermalization in nonintegrable systems~
\cite{Shenker2014,Roberts2015,Hosur2016,Kitaev2015,Bohrdt2017,Nahum2018,Keyserlingk2018}, 
 dynamics of systems with weak integrability breaking ~\cite{Polkovnikov2011,Marcuzzi2013}, artificial gauge field
~\cite{Dalibard2011}, topological band theory~\cite{Bansil2016}, and topological edge mode~\cite{SM} in the 
Su-Schrieffer-Heeger model~\cite{Su1979}.

\begin{acknowledgements} \textit{Acknowledgements.}---%
This work was partly supported by the Key-Area Research and Development Program
of GuangDong Province (Grant No. 2018B030326001), and the State Key Development
Program for Basic Research of China (Grants No. 2017YFA0304300).  Y. R. Z. was
supported by the Japan Society for the Promotion of Science (JSPS) (Postdoctoral
Fellowship via Grant No.~P19326, and  KAKENHI via Grant No.~JP19F19326). H. Y.
acknowledges support from the NSF of Beijing (Grant No.~Z190012), the NSFC of
China (Grants No.~11890704). H. F. acknowledges support from the National
Natural Science Foundation of China (Grant Nos.~11934018 and T2121001), Strategic Priority
Research Program of Chinese Academy of Sciences (Grant No.~XDB28000000), Beijing
Natural Science Foundation (Grant No.~Z200009).
\end{acknowledgements}

\clearpage \widetext
\begin{center}
	\section{Supplemental Material for \\ \textit{Probing Operator Spreading via Floquet Engineering in a Superconducting Circuit}}
\end{center}
%%%%%%%%%% Prefix a "S" to all equations, figures, tables and reset the counter %%%%%%%%%%
\setcounter{equation}{0} \setcounter{figure}{0}
\setcounter{table}{0} \setcounter{page}{1} \setcounter{secnumdepth}{3} \makeatletter
\renewcommand{\theequation}{S\arabic{equation}}
\renewcommand{\thefigure}{S\arabic{figure}}
\renewcommand{\bibnumfmt}[1]{[S#1]}
\renewcommand{\citenumfont}[1]{S#1}
%\renewcommand\thesection{S\arabic{section}}
%%%%%%%%%% Prefix a "S" to all equations, figures, tables and reset the counter %%%%%%%%%%

\makeatletter
\def\@hangfrom@section#1#2#3{\@hangfrom{#1#2#3}}
\makeatother

%---------------------------------------------------------------------------

\maketitle
This Supplemental Material contains four sections. In the first section, we
present further experimental details of this work, including device information,
measurement setup and method, readout calibration, single-qubit gate fidelity,
and Z pulse and phase calibrations. Then we discuss some nonideal aspects and
calculational details of our system, and their influence on the velocities of
single-photon quantum walk and operator spreading. The experimental observation
of the topologically protected edge mode in the Su-Schrieffer-Heeger model using
Floquet engineering will be demonstrated in the last section.

\section{Experimental Details}

\subsection{Device information} 
Our superconducting 10-qubit sample is shown in Fig.~\ref{fig_setup} at the
bottom (see also Fig.~1(a) in the main text), in which the qubits are transmons 
in the Xmon form~\cite{Koch2007,Barends2013}. The device was fabricated on a
$10\times10$ mm$^2$ sapphire substrate with a 430~$\mu$m thickness. An Al base
layer of 100~nm in thickness covering the whole substrate surface was first e-beam
evaporated and photolithographically patterned. Using the wet etching method,
the base-layer structures, including transmission lines, microwave coplanar
waveguide resonators, XY and Z control lines, and qubit capacitors were defined.
Josephson junctions in the dc-SQUID loops for tuning the critical current and
qubit frequency were fabricated by e-beam lithography and double-angle
evaporations. A 65~nm thick Al bottom electrode was evaporated at an angle of
$60^\circ$, and after oxidation in pure oxygen, a 100~nm thick Al counter
electrode was deposited at $0^\circ$. Finally, after lift-off,
airbridges~\cite{Chen2014} were fabricated to suppress parasitic modes as well
as the crosstalk among the control lines. Similar device fabrication processes
have been described elsewhere~\cite{Guo2021}.

In the 10-qubit chain, the nearby qubits are capacitively coupled with
approximately the same capacitance. The odd-number qubits have larger
capacitance due to an additional small cross-shaped area of the capacitor
electrodes (see Fig.~1(a) in the main text), so the maximum frequencies of the odd- and
even-number qubits stagger. The basic device parameters are listed in
Tab.~\ref{table_inf}. The frequency of the readout resonator $f_{\text{r}}$
ranges from 6.545 to 6.729~GHz, well located within the bandwidth of our
Josephson parametric amplifier (JPA). The maximum qubit frequency $f_{\text{m}}$
varies between 5.097 and 5.895~GHz, and $f_{\text{i}}$ is the qubit frequency at
idle point. The energy relaxation time $T_{1}$ and the dephasing time $T_2^*$
are both measured at the idle point. The coupling strength between two nearby
qubits is measured by vacuum Rabi oscillation at the working point. The
nearest-neighbor (NN) coupling strength $g_{j,j+1}$ is almost equal while the
next-nearest-neighbor (NNN) coupling strength $g_{j,j+2}$ of the odd- and even-number
qubits are different as a result of the different qubit capacitance. The
anharmonicity $U$ of odd- and even-number qubits are also different. In the
table, $F_{\text{g}}$ and $F_{\text{e}}$ are the readout fidelities of the
ground and first-excited states, respectively.

\subsection{Measurement setup and method} 
Our experimental setup is shown in Fig.~\ref{fig_setup}, in which from left to
right are the Z control (fast and dc), XY control, qubit readout (input and
output), and Josephson parametric amplifier (JPA) control lines, respectively.
The circuits include a series of attenuators, filters, dc-blocks, bias-Tees,
circulators, and amplifiers. The Z control lines provide dc and fast signals for
tuning the qubit frequency. The XY control lines send microwave pulses for the
manipulation of the qubit state. For the simultaneous state readout of all 10
qubits, ten-tone microwave pulse signals targeting the 10 readout
resonators can be sent through the transmission line and amplified successively
by the JPA, high electron mobility transistor (HEMT), and room-temperature
microwave amplifier, which are finally demodulated by the analog digital
converter.

The qubit readout resonator frequencies are displayed in Fig.~\ref{freq_space},
together with the qubit maximum frequencies and the frequencies at the idle and
working  points. In our experiment, the single-photon quantum walk, reversed
time evolution, and OTOCs are measured. For each measurement, all qubits are
first biased at their respective idle points for sufficiently long time during
which single gate operations are performed on certain qubits to prepare an
initial state of the system. Then all qubits are brought to their common working
point and Floquet driving is applied. The Floquet driving amplitude can be the
same (for single-photon quantum walk) or different (for reversed time evolution
and OTOC measurements) in this time evolution period or a Z-gate (or X-gate) operation 
is performed in the middle of the period (for OTOC measurements). At the end of this
period, all qubits are brought back to their idle points for the final-state
readout. The pulse sequences used for different experiments are presented in
Fig.~\ref{fig_sequence}, in which three main steps for the implementation of
the experiments, namely the initial-state preparation, Floquet driving and free
evolution, and state readout, can be seen.

\subsection{Readout calibration} 
The readout of the qubit state is performed at the qubit's idle point with the
duration time of demodulation 1.2~$\mu$s. Fig.~\ref{fig_readout}(a) shows the
gain profile of the JPA used in the experiment to improve the signal-to-noise
ratio for the state readout. In our experiment, the idle point of each qubit is
chosen in the vicinity of the qubit sweet point in order to have a longer
dephasing time, which is also well away from the frequencies of other unwanted
two-level systems to avoid their couplings. Meanwhile, the frequency difference
between the nearest-neighbor qubits is set as large as possible to reduce their
XY crosstalk. For the present study, simultaneous readout of all 10 qubits in
the chain is required. In Fig.~\ref{fig_readout}(b), we show the IQ data of such
joint readout of the qubit states with high fidelities (see
Tab.~\ref{table_inf}). In order to correct the errors from initial state
preparation and decoherence during measurement, we use the calibration matrix to
reconstruct the readout results, defined as
\begin{equation}
	T_{j}=
	\left(
	\begin{array}{cc}
		F_{g,j} & 1-F_{e,j}\\
		1-F_{g,j} & F_{e,j}\\
	\end{array}
	\right)~,
	\label{readout cali}
\end{equation}
where $F_{g,j}$ and  $F_{e,j}$ are the readout fidelities of the $j$-th qubit
after being well initialized in the ground and first-excited states,
respectively (see Tab.~\ref{table_inf}). With the calibration matrix, the
correct probability of different qubit states can be found~\cite{Yan2019}.
Finally, in the experiment of OTOCs, the post selection is applied for the
readout results based on the conservation of photon number and the constraint of
the qubit computational subspace.

\subsection{Single-qubit gate fidelity} 
We use $X_{\pi}$ gate for the preparation of the excited state $\ket{1}$ for the
selected qubits. Each $X_{\pi}$ gate contains two $X_{\pi/2}$ gates, which are
accomplished by Gaussian-enveloped sinusoidal pulses with the length  of 20~ns.
The corresponding amplitude is optimized by repeating multiple sequences. The
qubit frequencies are accurately determined by Ramsey interferometry
measurement. In order to eliminate the phase error induced by the XY drive, we
use AllXY sequence to optimize the DRAG factor~\cite{Reed2013}. Using the
randomized benchmarking method (see Fig.~\ref{fig_rb} for the result of
$Q_2$ as an example), we obtain the single-qubit gate fidelities, some
of which are listed in Tab.~\ref{table2}.

\subsection{Z pulse calibration and correction} 
In our experiment, various Z control line signals are required for different purposes
for setting fixed qubit frequencies, performing gate operations, and producing a
cosine-form oscillation for the qubit frequency with $\nu$=120~MHz.
Below we describe the calibration of the Z pulse crosstalk,
Z pulse shape correction, and optimized Z-gate operation, which are necessary for
the precise qubit manipulations in the present experiment.

Due to the existing mutual inductance among the qubit control lines, the bias
current applied on one qubit will also affect other qubits. In order to
calibrate this Z pulse crosstalk, we first use dc bias to set a qubit $Q_i$ at
a point in the energy spectrum that is sensitive to the flux, and excite it
with a resonant microwave for 19~$\mu$s. At the same time, we scan the Z pulse
amplitudes for $Q_i$ and another qubit $Q_j$ with the same pulse length of
20~$\mu$s. Since the qubit frequency changes linearly with respect to small
offset, we expect that the position of the resonance peak also changes linearly
with these two offsets. Hence the crosstalk matrix element for these two qubits
can be found from the slope of the resonance peak plotted against the two
offsets
\begin{equation}
	c_{ij} = -\frac{\Delta f_{i}}{\Delta f_{j}},
	\label{crosstalk}
\end{equation}
where $\Delta f_i$ is the frequency offset of $Q_i$. The measured
crosstalk matrix is shown in Fig.~\ref{Zpulse_Cali}(a).

\begin{figure*}[tbh]
			\centering
			\includegraphics[scale=0.42]{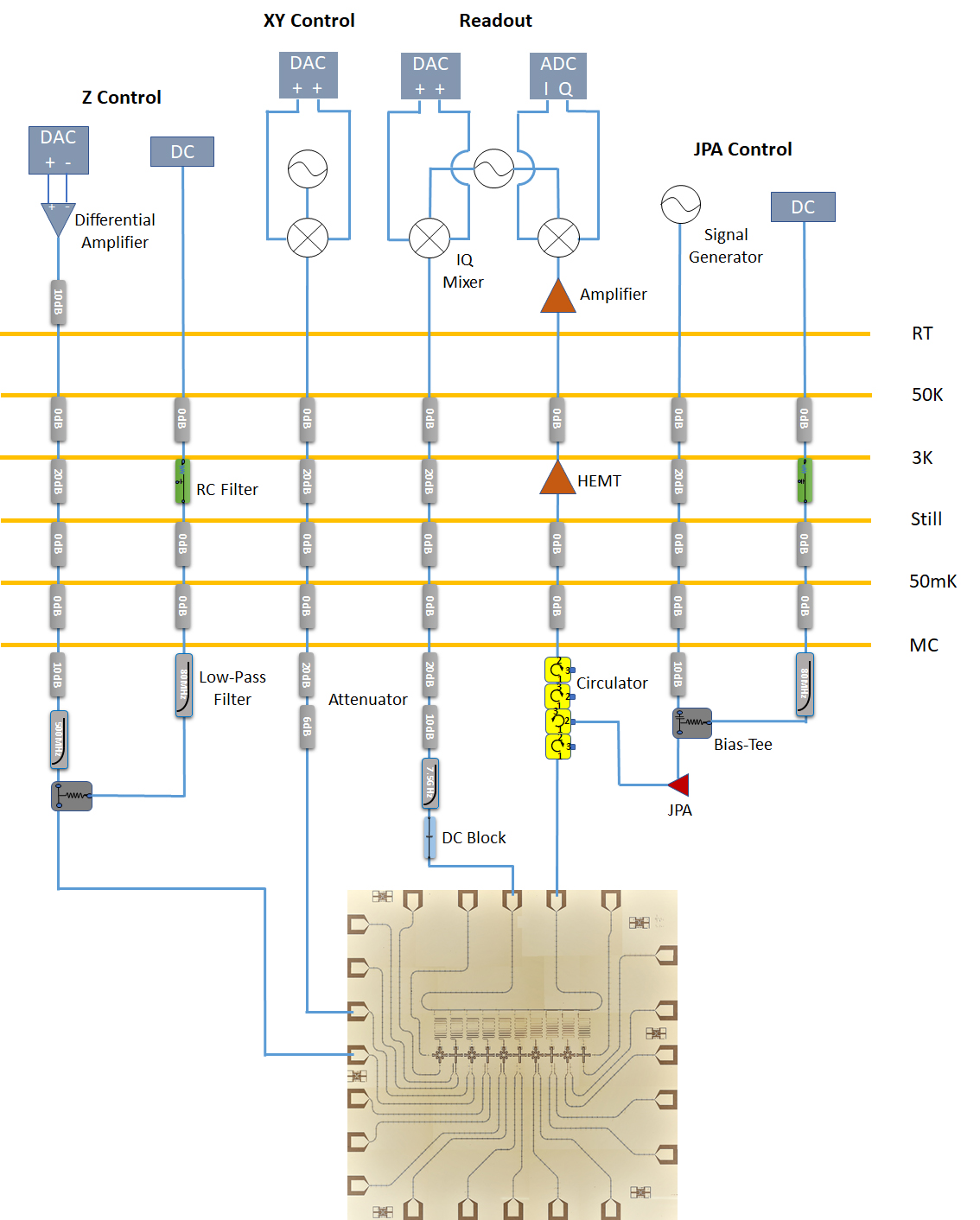}  
			\caption{Experimental setup. From left to right are the Z control (fast and
				dc), XY control, qubit readout (input and output), and JPA control lines,
				respectively. The XY and Z control lines are shown only for the third qubit
				$Q_3$ for simplicity.}
			\label{fig_setup}
		\end{figure*}

As is shown in Fig.~\ref{fig_setup}, starting from the DAC outputs, our Z
control lines consist of differential amplifiers, bias-Tees, filters, and
attenuators. Their limited bandwidth distorts the shape of Z pulse signals,
which will in turn accumulate the unwanted phase error, as can be seen in the
left panel of Fig.~\ref{Zpulse_Cali}(b). We use the deconvolution method to
correct the distortion. By scanning the shape of the Z pulse with
$X_{\pi}$ gate, we can fit the peak to get the response function with respect to
the control system. We predistort the shape of Z pulse on the basis of the
response function for the correction~\cite{Guo2021}. In the right panel of
Fig.~\ref{Zpulse_Cali}(b), we show the shape of the Z pulse, which becomes flat,
after the correction.

In the measurement of OTOCs, we need to perform a Z gate (phase-flip gate) or an
X gate (bit-flip gate) operation on the tenth qubit $Q_{10}$, which is done at
the idle point. At the same time, the phase accumulation for the other qubits
during the operation should be avoided. This is realized by applying a Z gate to
$Q_{10}$ and identity gates to the other qubits by Z pulses having the same
duration as that of the Z gate pulse. To calibrate the different Z pulse
amplitudes, we insert a Z pulse between two $X_{\frac{\pi}{2}}$ gates, change
the amplitude (voltage) of the Z pulse, and readout the qubit population. The
voltages of the Z pulse corresponding to the populations of 0 and 1 in the
oscillating dependence reflect the Z pulse amplitudes required for the Z gate
and identity, respectively. For more accurate determination, we insert an
odd-number of identical Z pulses and readout the population again. The results
for $Q_{1}$ are shown in Fig.~\ref{Zpulse_Cali}(c) as an example. We choose the
voltage indicated by the white (red) line as the parameter for the Z (identity)
gate.

\subsection{Calibration and correction of initial state phase and dynamical phase}

\begin{figure*}[t]
			\centering 
			
			\includegraphics[scale=0.7]{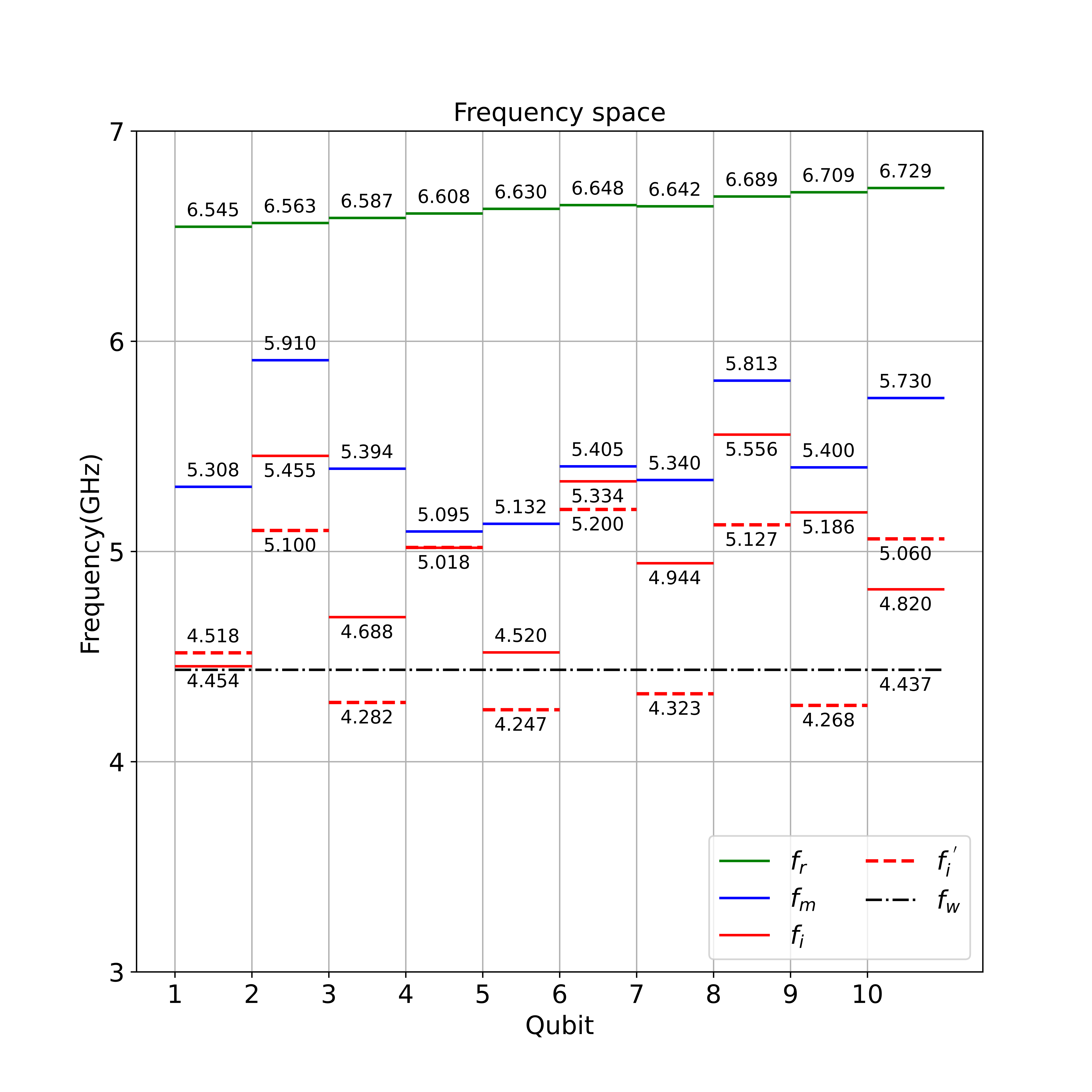}  %这个是图片的绝对路径
			
			\caption{Device frequency diagram. The green, blue, and red lines represent the
				readout resonator frequency $f_r$, qubit maximum frequency $f_m$, and qubit
				idle frequency $f_{i}$, respectively.The red dashed lines represent the idle frequency  
				used only for the measurement of $C_j^{\text{XX}}$. The frequency of the qubit
				working point($f_w$) is shown with a long black dash-dotted line.
			}
			
			\label{freq_space}
		\end{figure*}

\begin{figure*}[t]     		
			\includegraphics[width=1\textwidth]{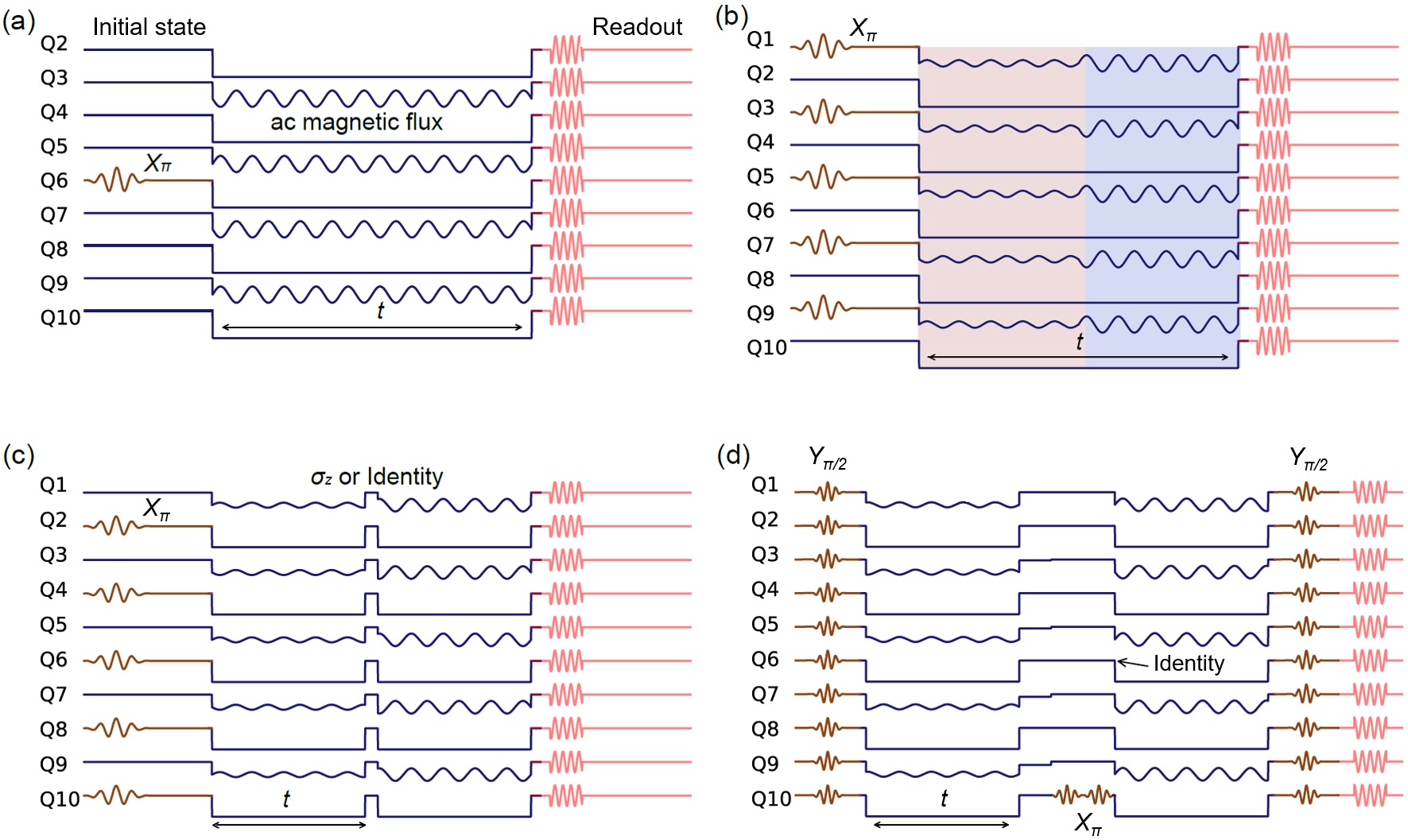} \caption{Pulse sequences for
				the experiments of (a) single-photon quantum walk, (b) reversed time evolution,
				(c) ZZ OTOC, and (d) XX OTOC. The orange, blue, and red pulses represent XY, Z,
				and readout drives, respectively.}
			\label{fig_sequence}
		\end{figure*}

\begin{figure*}[tbh]
			\centering
			\includegraphics[scale=0.65]{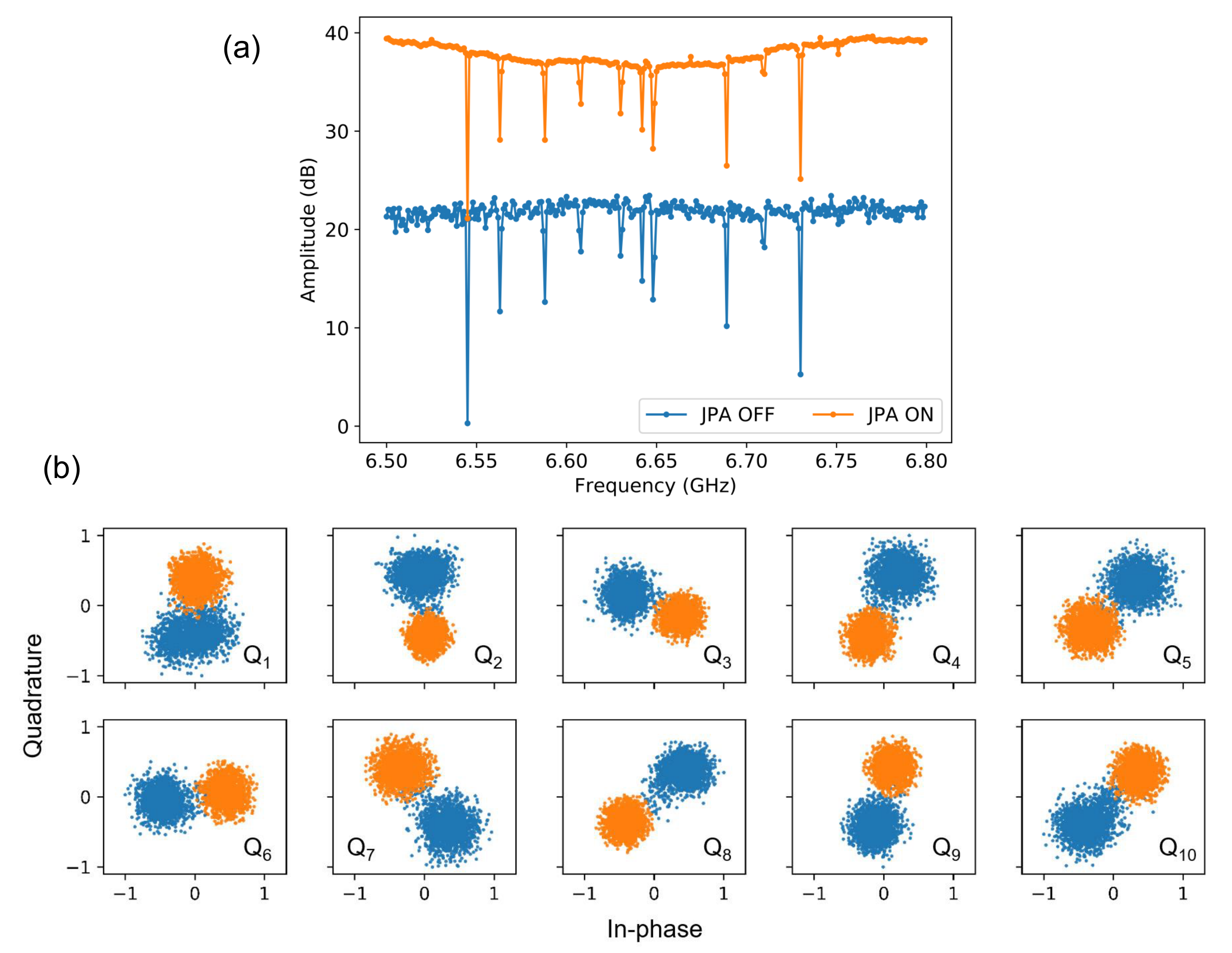} 
			\caption{Qubit-state readout. (a) The transmission spectra of the readout
				signals with JPA on (blue) and off (orange). (b) IQ data of the readout signals
				of the ground (blue) and first-excited (orange) states, where 10 qubits are
				measured simultaneously.}
			\label{fig_readout}
		\end{figure*}

\begin{figure*}[tbh]
			\centering
			\includegraphics[scale=0.55]{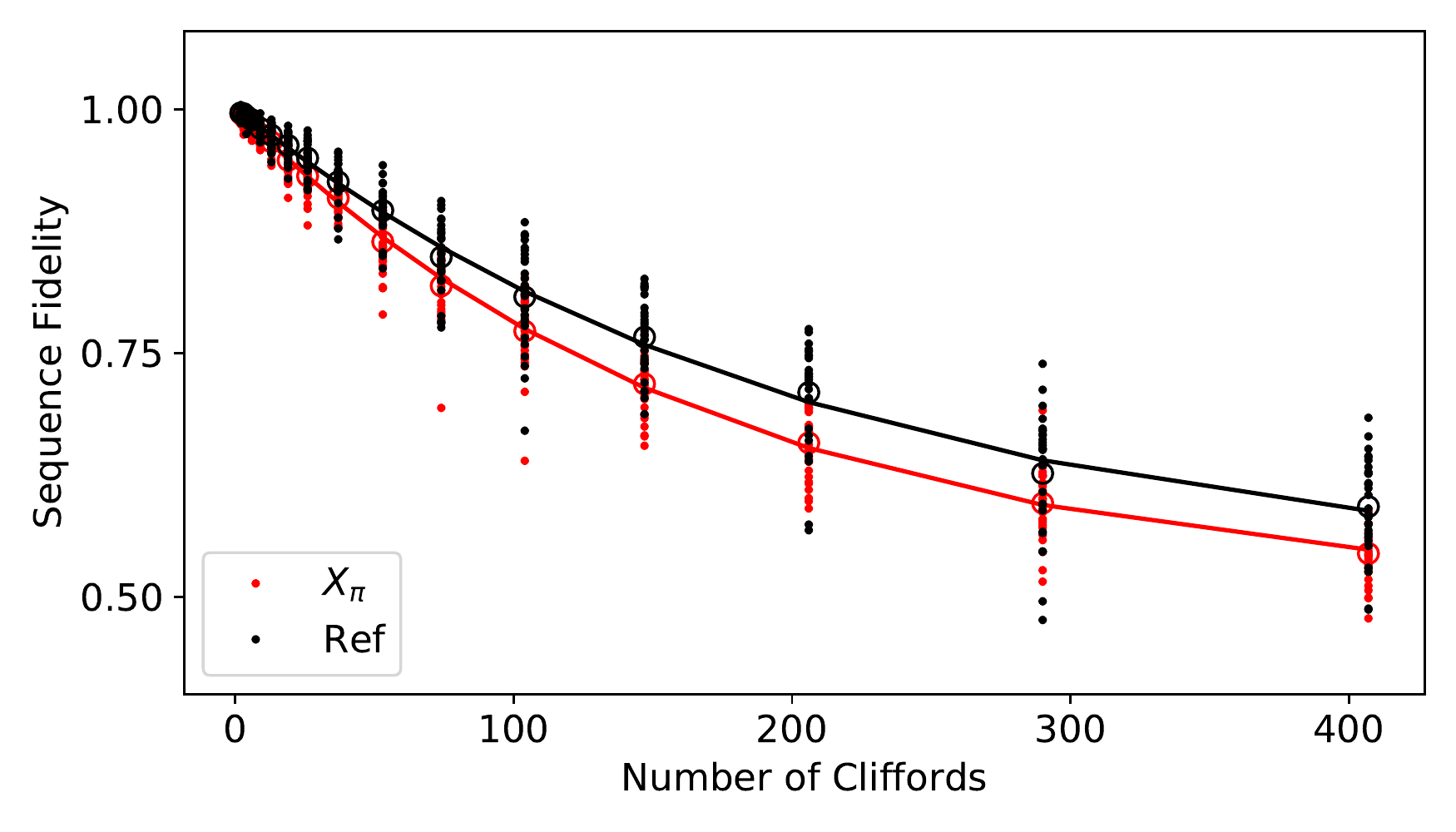}  
			\caption{Randomized benchmarking result for $X_{\pi}$ gate measured on $Q_2$
				with a gate fidelity of 99.93$\%$.} 
			\label{fig_rb}
		\end{figure*}

\begin{figure*}[tbh]
			\centering
			\includegraphics[scale=0.65]{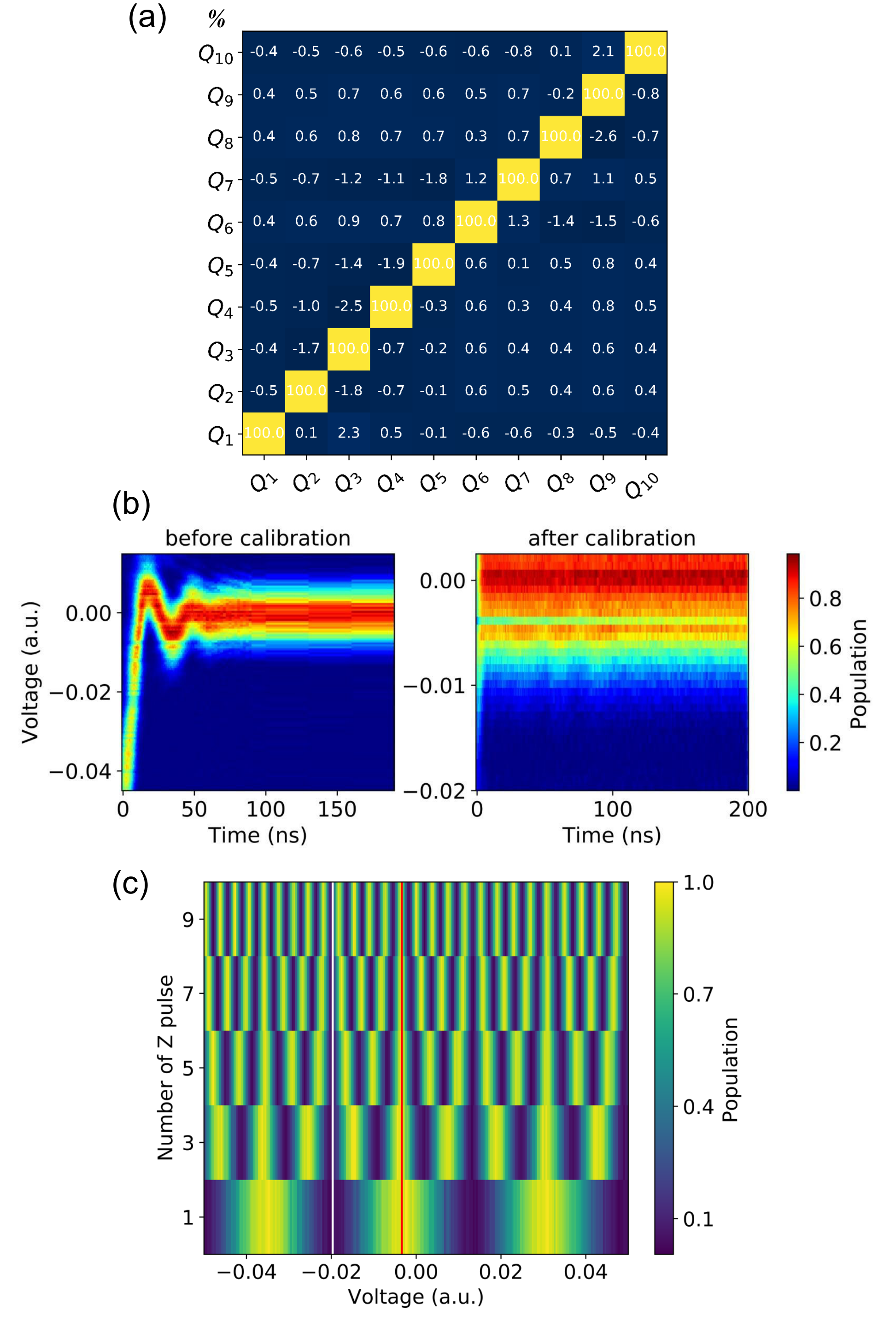}  
			\caption{Z pulse calibration and correction. (a) Z pulse crosstalk matrix of
				the 10-qubit chain. (b) The original distorted Z pulse (left panel) and the one
				after correction (right panel). (c) Measured populations of $Q_1$ as functions
				of the amplitude and the number of Z pulses inserted between two $X_{\pi /2}$
				gates. We choose the voltages indicated by the white (red) line as the pulse
				amplitude for the Z (identity) gate.}
			\label{Zpulse_Cali}
		\end{figure*}

\begin{figure*}[t]
			\centering	
			\includegraphics[scale=0.6]{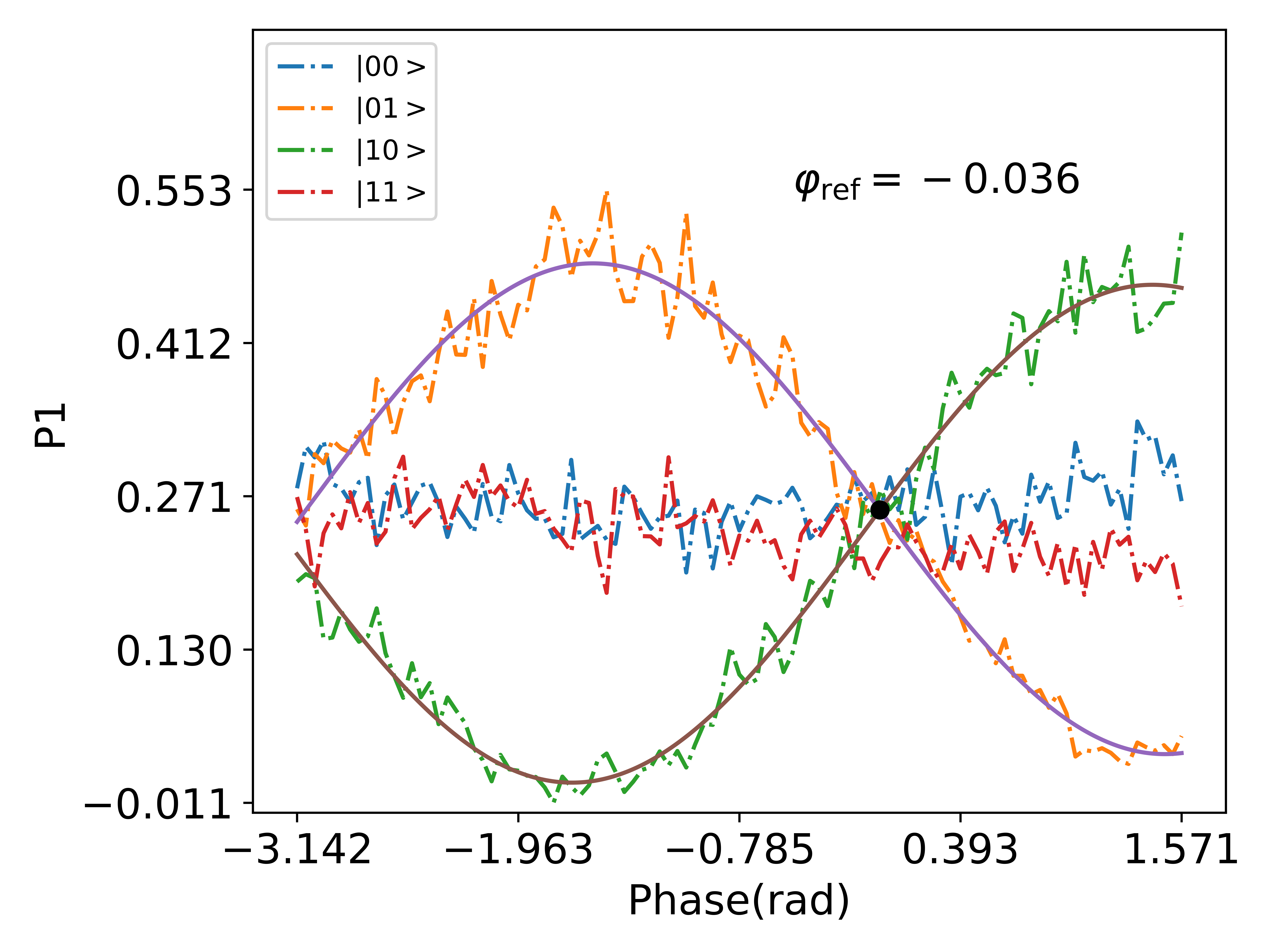}  %这个是图片的绝对路径
			\caption{Calibration of the phase of the initial state. Curves with different
				colors label  probabilities of four two-qubit computational basis. The solid
				lines are fits with cosine function. The black dot indicates the final
				reference phase $\varphi_{\rm ref}$ = -0.036.}
			
			\label{fig12}
		\end{figure*}

In the preparation of the initial state $\ket{++++++++++}$ for the measurement
of $C_j^{\text{XX}}$, we use a set of idle points ($f'_i$) shown as dashed lines
in Fig.~\ref{freq_space}, with larger separations between nearby qubits to
reduce the effect of ZZ interaction, which is evaluated by Ramsey interference
experiment~\cite{Ku2020}. The 10-qubit product state is prepared by
$Y_{\frac{\pi}{2}}$ gates, see Fig.~\ref{fig_sequence}(d), where
$\ket{+}=\frac{1}{\sqrt{2}}(\ket{0}+\ket{1})$. To make the $x$-axis of each
Bloch sphere align, we calibrate and correct the phase of each XY driving with
respect to the rotating frame of frequency $f'_i$~\cite{Song2017}. For instance,
we prepare the initial state of $Q_1$ and $Q_2$ as
$\frac{1}{\sqrt{2}}(\ket{0}+e^{i\varphi_{\rm ref}}\ket{1}) \otimes
\frac{1}{\sqrt{2}}(\ket{0}+e^{i(\varphi+\varphi_0)}\ket{1})$, where $\varphi_0$
originates from the phase difference between the two XY drivings. We choose
$Q_1$ as the reference and fix $\varphi_{\rm ref}=0$. Then we set the two qubits
at resonance at the working point $f_w$ for time $t=\frac{\pi}{4g_{j,j+1}}$. The
two-qubit state will become $\frac{1}{2}\ket{00} +
\frac{e^{i(\varphi+\varphi_0)}+i}{2\sqrt{2}}\ket{01} +
\frac{ie^{i(\varphi+\varphi_0)}+1}{2\sqrt{2}}\ket{10} +
\frac{e^{i(\varphi+\varphi_0)}}{2}\ket{11}$. We adjust the phase $\varphi$ to
have $\varphi+\varphi_0=0$, which gives equal probabilities of the four
two-qubit computational basis. Finally, we choose $-\varphi_0$ as new
$\varphi_{\rm ref}$ (black dot in Fig.~\ref{fig12}) for correcting the phase
between $Q_2$ and $Q_3$. The procedure is repeated pairwisely and successively
to correct  phases for all qubits in the initial state.

When the qubit frequency is biased away from the idle point by a Z pulse in the
experiment, it deviates from the microwave frame and accumulates extra dynamical
phase. We perform Ramsey-like measurement to determine the dynamical phase and
make the compensation \cite{Guo2021}.

\section{Velocities of single-photon quantum walk and operator spreading}

In the main text, we have mentioned that the group velocity of single-photon
quantum walk is $\varepsilon$-dependent and equals the velocity of the operator
spreading seen from $C_j^{\text{ZZ}}$ with the same coupling strength. Here we
present the details for the evaluations of both velocities together with their
error analysis.

\begin{figure*}[t]     		
			\includegraphics[width=1\textwidth]{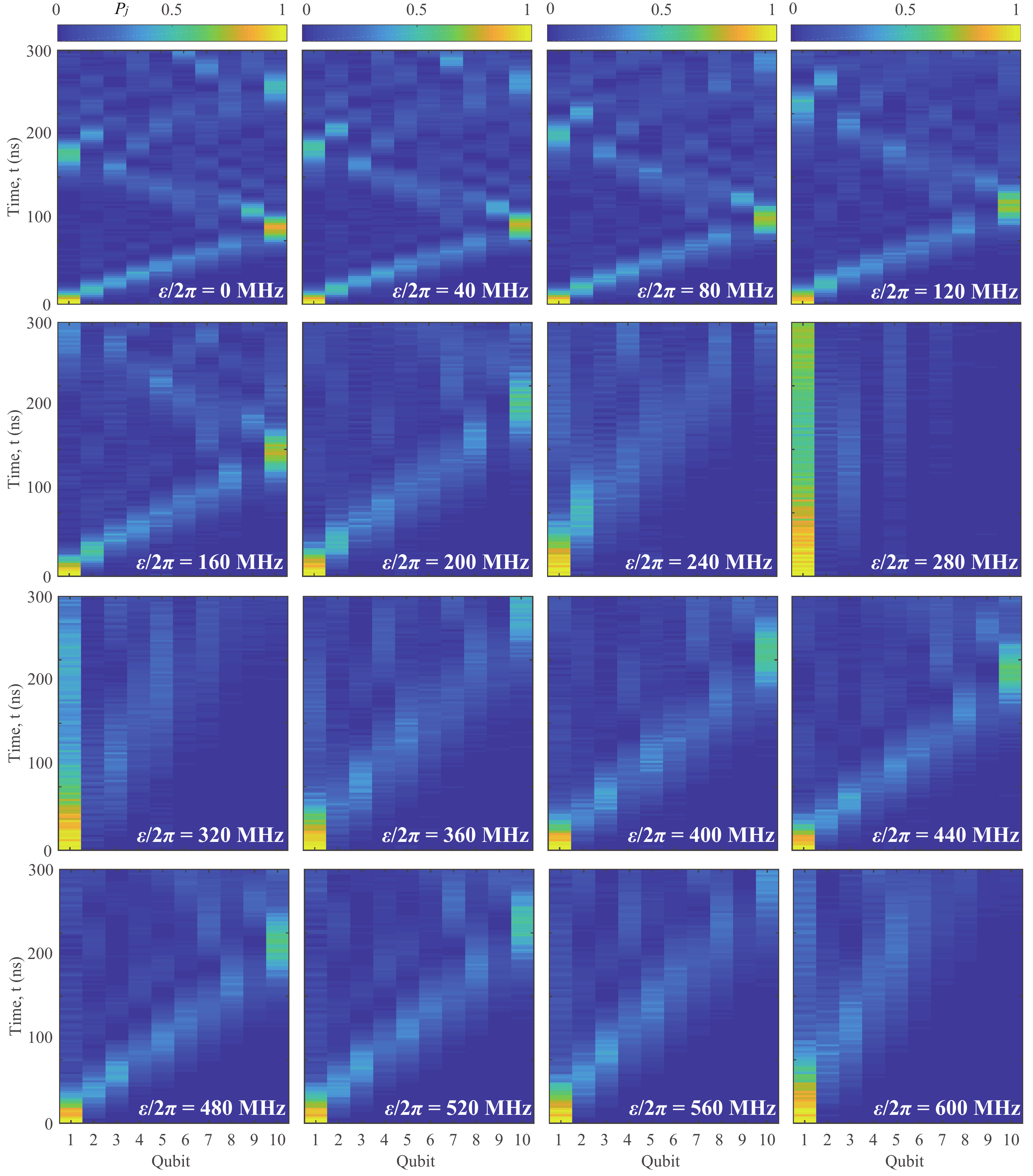}
			\caption{Experimental results of single-photon quantum walk of the 10-qubit
				chain under periodic driving with different $\varepsilon$. The initial state is
				$\ket{1000000000}$.  Each point is the average result of $8,000$ single-shot
				readouts.}
			\label{fig_lc}
		\end{figure*}

\begin{figure*}[t]     		
			\includegraphics[width=0.8\textwidth]{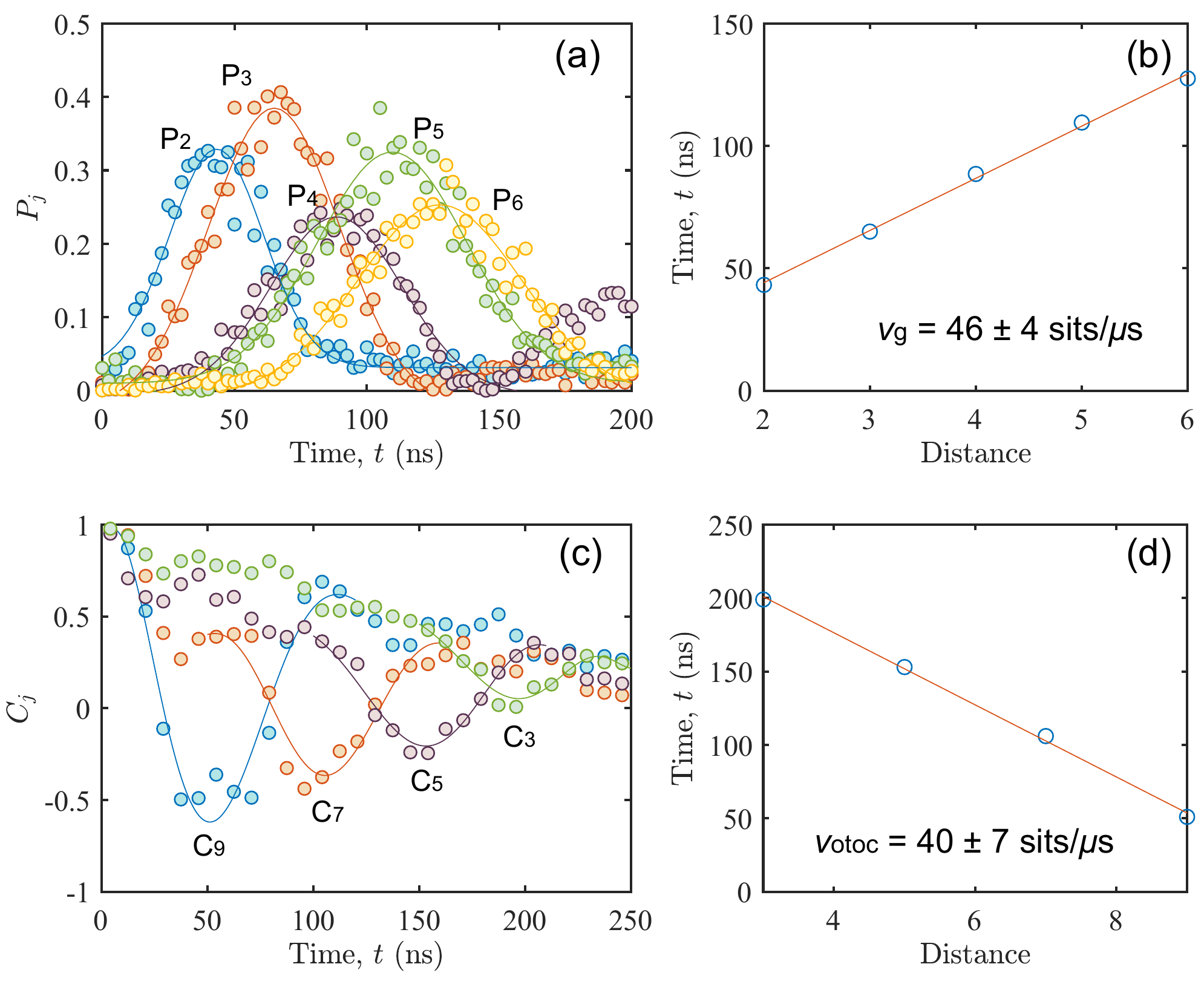}
			\caption{Fitting velocities of (a, b) single-photon quantum walk with
				$\varepsilon/2\pi=400$~MHz and corresponding coupling strength of 4 MHz, and
				(c,d) operator spreading with the same coupling strengths in absolute values.}
			\label{v_fit}
		\end{figure*}

In Fig.~\ref{fig_lc}, we show the experimental results of single-photon quantum
walks in the 10-qubit chain with varying $\varepsilon$ and an initial state of
$\ket{1000000000}$. To obtain the group velocities~\cite{Yan2019,Ye2019}, we
perform Gaussian fit to the data of $P_2(t)$--$P_6(t)$ and find the
corresponding times of the first front of the fitting curves. Through a linear
fit between the front time and qubit number, the group velocities can be
obtained. In Fig.~\ref{v_fit}(a--b), we show an example for the result with
$\varepsilon/2\pi=400$~MHz, for which the group velocity is calculated to be
$v_g=46\pm4$~sites/$\mu$s. For the experimental results of $C_j^{\text{ZZ}}$,
the Gaussian fit is not applicable and the polynomial  fit is used.
Fig.~\ref{v_fit}(c) shows the experimental and fitting $C_3(t)$, $C_5(t)$,
$C_7(t)$, $C_9(t)$. The front time of $C_j$ is plotted against qubit number in
Fig.~\ref{v_fit}(d), from which the operator spreading velocity of
$v_{\text{otoc}}=40\pm7$~sites/$\mu$s is found.

For an isotropic $XY$ model in Eq.~(3) with only NN coupling, one expects that
the dynamics of the system can be perfectly reversed. Now we consider the
influence of the existing nonideal NNN coupling and the high-level occupation
that leads the system away from the ideal isotropic $XY$ model. We discuss
how these two limitations affect the velocities of single-photon  quantum walk
and operator spreading seen from $C_j^{\text{ZZ}}$.

Tab.~\ref{table_inf} shows that the NNN coupling strengths are about $1\times
2\pi$~MHz and $0.5\times 2\pi$~MHz for the odd- and even-site qubits,
respectively. In order to partly suppress the NNN coupling, we apply ac magnetic
flux on odd qubits with staggered phase, i.e., $\varepsilon_1,
\varepsilon_5,\varepsilon_9=\varepsilon$ while
$\varepsilon_3,\varepsilon_7=-\varepsilon$. In this case, the effective coupling
strength for the odd qubits becomes $\mathcal{J}_0(2\varepsilon/\nu)\times
2\pi$~MHz, while it does not change for the even qubits. This results in the
reduced NNN coupling strength for the odd qubits. For instance, we have
$\mathcal{J}_0(2\varepsilon/\nu)\approx -0.388$ for $\varepsilon/2\pi=213.6$~MHz
and  $\mathcal{J}_0(2\varepsilon/\nu)\approx 0.282$ for
$\varepsilon/2\pi=400$~MHz, so the NNN coupling strength for the odd qubits
becomes smaller compared to the original values.

For the discussion of $C_j^{\text{ZZ}}$ and operator spreading, we consider 
the qubit two-level subspace and take into account the influence of the
second-excited state via the so-called ZZ interaction~\cite{McKay2017,Braum2021}.
Experimentally the post selection is applied for the measurement of OTOCs, where
the high-level occupations of the final states are neglected. The Hamiltonian
considering the ZZ interaction reads
\begin{align} \nonumber
	&\hat H^{\prime}=\hat H_{\text{eff}}+ \epsilon \hat H_{\text{zz}}~,\\
	&\hat H_{\text{zz}} =-\frac{1}{2}\sum_{j=1}^{9}(1+\hat{\sigma}^z_j\hat{\sigma}^z_{j+1})~,
\end{align}
where $\epsilon \sim g^2_{\text{eff}}/U \sim 0.065 \times 2\pi$~MHz.

In Tab.~\ref{table_v}, we compare the experimental velocities of single-photon
quantum walk and operator spreading with the numerical ones with or without
considering the NNN coupling and ZZ interaction.  We can find that NNN coupling
tends to enlarge the velocity of single-photon quantum walk, while the velocity
of operator spreading is robust to the NNN coupling and ZZ interaction. In the
last column of the table, a homogeneous XY chain with 25 spins and common NN
coupling strength of $g$ MHz are considered. In this case, the two velocities
are  identical.

\section{Numerical Calculation of $C_j^{\text{XX}}$}

To explain experimental results of $C_j^{\text{XX}}$, we find that the qubit
third level must be considered in numerical calculations during the
entire dynamical process. Here, we replace the Pauli 
operator $\hat{\sigma}^x_j$ with
\begin{equation}
	\hat{\Sigma}^x_{j}=
	\left(
	\begin{array}{ccc}
		0 & 1 & 0\\
		1 & 0 & 0\\
		0 & 0 & \eta\\
	\end{array}
	\right)~
	\label{sigmax3a}
\end{equation}
for the butterfly operator ($\eta$=1) and the measurement operator ($\eta$=0) in
Eq.~(6) in the main text. This implies that we neglect the third-level influence
only in the readout. This treatment provides a reasonable
description for the experimental $C_j^{\text{XX}}$. 

We also present the numerical result of $C_j^{\text{XX}}$ in a larger time scale 
without considering the qubit third level, see Fig. S10. The result without using Floquet driving or  considering NNN coupling is also shown for comparison. These results demonstrate a nearly periodic behavior of $C_j^{\text{XX}}$ for the near integrable system and the blurring of propagation features arising from Floquet driving and NNN coupling.
%by considering two levels of each qubit, see Fig.~\ref{otocxx}.
%We also present numerical simulation without
%using Floquet drivings and considering NNN couplings.
%Form these results, we can find that there is indeed a nearly periodic behavior of $C_j^{\text{XX}}$ in a larger time scale
%for this near integrable system.
%We find that similar
%consideration also gives a satisfactory fit for the experimental data of
%$C_j^{\text{ZZ}}$ as the ZZ-interaction treatment in Eq.~(\ref{Hxy}).

\section{Su-Schrieffer-Heeger model and topologically protected edge mode} 
In the main text, we have used Floquet engineering to realize the homogeneous NN
coupling for different studies. Here we demonstrate that the spatially periodic
coupling strength in the qubit chain can be realized using the method for the
study of the spin Su-Schrieffer-Heeger 1D compound lattice model~\cite{Su1979}
and its topological properties. To this end, we divide the 10-qubit chain into two
sublattices labelled by $A$ and $B$, of which the Hamiltonian reads
\begin{align} 
	\hat H_{\text{SSH}} =\sum_{j}(g_{o}^{\text{eff}}\hat{\sigma}^+_{A,j} \hat{\sigma}^-_{B,j}+g_{e}^{\text{eff}}\hat{\sigma}^+_{B,j} \hat{\sigma}^-_{A,j+1} + \text{H.c.}),
\end{align}
where we label the odd-site qubit $Q_{2j-1}$ as site $(A,j)$ and even-site qubit
$Q_{2j}$ as site $(B,j)$. We find that when the effective intracell coupling
strength is larger than  intercell coupling, i.e.,
$|g_{o}^{\text{eff}}|>|g_{e}^{\text{eff}}|$, the system is topologically
trivial. On the other hand, when
$|g_{o}^{\text{eff}}|<|g_{e}^{\text{eff}}|$, the system becomes topologically
nontrivial with the emergence of topological edge modes.

Experimentally we Floquet drive $Q_3,\ Q_4,\ Q_7$, and $Q_8$
with the amplitude $\varepsilon/2\pi=156$~MHz, so the effective intracell
coupling $g_{o}^{\text{eff}}=g_{2j-1,2j}$, and intercell coupling
$g_{e}^{\text{eff}}\approx 0.5 g_{2j,2j+1}<g_{o}^{\text{eff}}$. In
Fig.~\ref{fig_ssh}(a), we present the measured quench dynamics of single
excitation in this case. We can see that the single excitation is delocalized
when initially placed at the left-most qubit $Q_1$, so the system is
topologically trivial. Figure~\ref{fig_ssh}(b) shows the case when ac magnetic
flux is applied on $Q_2,\ Q_3,\ Q_6$, $Q_7$, and $Q_9$ with the amplitude
$\varepsilon/2\pi=156$~MHz. The system becomes topologically nontrivial in this
case since the single excitation initially placed at the edge remains localized,
which is a strong dynamical signature of the existence of topologically
protected edge mode.

\begin{center}
	\title{Supplemental Tables}
\end{center}

\begin{table*}[tbh]
	\vspace{15pt}
	\setlength{\tabcolsep}{7pt}
	\centering
	\resizebox{\textwidth}{!}{
		\begin{tabular}{c |c c c c c c c c c c  }
			\hline
			\hline
			& $Q_1$ & $Q_2$ & $Q_3$ & $Q_4$ & $Q_5$ & $Q_6$ & $Q_7$ & $Q_8$ & $Q_9$ & $Q_{10}$ \\
			\hline
			$f_{\text{r}}$ (GHz)  & 6.545 & 6.563 & 6.587 & 6.609 & 6.630 & 6.648 & 6.642 & 6.689 & 6.709 & 6.729 \\
			$f_{\text{m}}$ (GHz) & 5.308 & 5.896 & 5.387 & 5.097 & 5.144 & 5.395 & 5.327 & 5.804 & 5.389 & 5.726 \\
			$f_{\text{i}}$ (GHz) & 4.454 & 5.455 & 4.688 & 5.018 & 4.520 & 5.334 & 4.944 & 5.556 & 5.186 & 4.820 \\
			$f'_{\text{i}}$ (GHz) & 4.518 & 5.100 & 4.282 & 5.018 & 4.247 & 5.200 & 4.323 & 5.127 & 4.268 & 5.060 \\
			$U/2\pi$ (MHz)  & -212 & -264 & -210 & -268 & -212 & -268 & -214 & -264 & -214 & -264 \\
			$T_{1}$ ($\mu$s)  & 47.6 & 26.3 & 40.4 & 17.5 & 23 & 29.3 & 46 & 30.8 & 38.2 & 29.3 \\
			$T_{2}^*$ ($\mu$s) & 1.97 & 2.34 & 1.74 & 6.36 & 2.12 & 36.7 & 2.63 & 3.79 & 3.51 & 1.54 \\
			$F_{\text{g}}$ (\%) & 97.2 & 99.2 & 99.2 & 99.7 & 98.1 & 99.4 & 99.3 & 99.4 & 99.3 & 99.3 \\
			$F_{\text{e}}$ (\%)  & 90.4 & 92.6 & 92.4 & 90.7 & 90.3 & 91.7 & 89.9 & 91.7 & 92.5 & 85.6 \\
			\hline
			$g_{j,j+1}/2\pi$ (MHz) & \multicolumn{10}{c} {10.72 \hspace{3.25mm} 10.73 \hspace{3.25mm} 10.99 \hspace{3.25mm} 11.05 
				\hspace{3.25mm} 10.88 \hspace{3.25mm} 10.48 \hspace{3.25mm} 10.86 \hspace{3.25mm} 10.79 \hspace{3.25mm} 10.78}\\
			%     	   $g_{j,j+1}/2\pi$ (MHz) & 10.72 & 10.73 & 10.99 & 11.05 & 10.88 & 10.48 & 10.86 & 10.79 & 10.78&\\
			\hline
			$g_{j,j+2}/2\pi$ (MHz)  && 0.98 & 0.49 & 0.96 & 0.49 & 0.96 & 0.49 & 0.97 & 0.48&\\
			\hline
			\hline
	\end{tabular}}
	\caption{Basic device parameters. $f_{\text{r}}$ is the readout resonator
		frequency, $f_{\text{m}}$ is the qubit maximum frequency, and $f_{\text{i}}$ is
		the qubit idle frequency ($f'_{\text{i}}$ is the idle frequency used only for 
		the measurement of $C_j^{\text{XX}}$). $U$ is the qubit anharmonicity. $T_{1}$
		and $T_2^*$ are the energy relaxation time and dephasing time of the qubit at
		idle point. $F_{\text{g}}$ and $F_{\text{e}}$  are the readout fidelities for
		the ground and first-excited states, respectively. $g_{j,j+1}$ and $g_{j,j+2}$
		are the coupling strengths of nearest-neighbor (NN) and next-nearest-neighbor
		(NNN) qubits, respectively.}
	\label{table_inf}
\end{table*}

\begin{table*}[t]
	\vspace{5pt}
	\setlength{\tabcolsep}{7pt}
	\centering
	\resizebox{\textwidth}{!}{
		% \begin{threeparttable}
			\begin{tabular}{c |c c c c c c c c c c  }
				\hline
				\hline
				& $Q_1$ & $Q_2$ & $Q_3$ & $Q_4$ & $Q_5$ & $Q_6$ & $Q_7$ & $Q_8$ & $Q_9$ & $Q_{10}$ \\
				\hline
				$X_{\pi}(\%)$  & 99.81 & 99.93 & 99.91 & 99.89 & 99.84 & 99.92 & 99.76 & 99.96 & 99.88 & 99.59 \\
				$X_{\frac{\pi}{2}}(\%)$  & 99.91 & 99.95 & 99.89 & 99.97 & 99.87 & 99.98 & 99.40 & 99.87 & 99.71 & 99.81 \\
				$Y_{\frac{\pi}{2}}(\%)$ & 99.89 & 99.95 & 99.90 & 99.93 & 99.89 & 99.97 & 99.24 & 99.77 & 99.77 & 99.79 \\
				\hline
				\hline
		\end{tabular} }
		\caption{Fidelities of single-qubit gates measured by the randomized
			benchmarking method. $X_{\theta}$ ($Y_{\theta}$) represents a rotation around
			the $x$- ($y$-) axis for an angle of $\theta$.}
		\label{table2}
	\end{table*}
	
	\begin{table*}[t]
		\vspace{5pt}
		\setlength{\tabcolsep}{7pt}
		\centering
		\resizebox{\textwidth}{!}{
			% \begin{threeparttable}
				\begin{tabular}{c |c c c c  c  }
					\hline
					\hline
					Velocity(sites/$\mu$s)&Exp. &  NNN coupling  & ZZ coupling  & No NNN or ZZ coupling & Homogeneous  \\
					\hline
					Quantum walk  & $46\pm4$    & $50\pm5$      & $42\pm2$    & $42\pm2$    & $(1.85\pm0.02)g$  \\
					Operator spreading      & $40\pm7$    & $39\pm9$       & $39\pm10$   & $39\pm10$   & $(1.85\pm0.02)g$  \\
					\hline
					\hline
			\end{tabular} }
			\caption{Comparison of the velocities of single-photon quantum walk and operator spreading.
				In the last column, a  homogeneous $XY$ chain with 25 spins and an NN coupling
				of $g_j=g$~MHz are considered.}
			\label{table_v}
		\end{table*}

		%\begin{figure*}[tbh]
		%	\centering
		%	\includegraphics[scale=0.65]{freq_space.pdf}  
		%	\caption{Device frequency diagram.  The green, blue, and red lines represent
			%	the readout resonator frequency $f_{\text{r}}$,  qubit maximum frequency
			%	$f_{\text{m}}$, and qubit idle frequency $f_{\text{i}}$, respectively. The
			%	frequency of the qubit working point is shown with a long black dashed line.}
		%	\label{freq_space}
		%\end{figure*}
		
		
		\begin{figure*}[t]     		
			\includegraphics[width=0.8\textwidth]{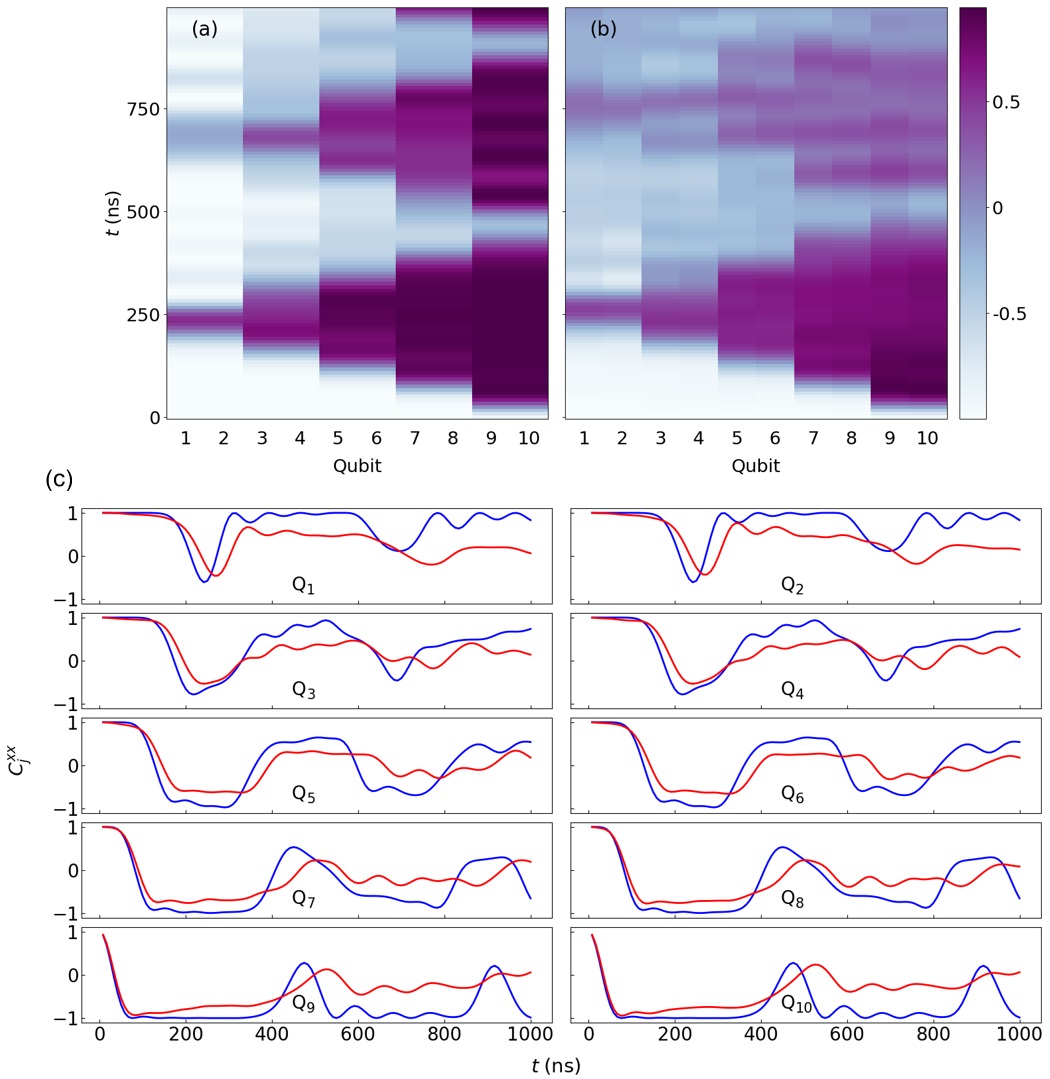} \caption{Numerical results of
				$C_j^{\text{XX}}$ by considering two levels. (a) Homogeneous NN coupling of
				4 MHz without NNN coupling are considered, and time reversal is realized
				by directly changing the sign of NN coupling without Floquet driving. (b) With
				Floquet driving and using experimental parameters as described in the main
				text. The blue and red lines in (c) are the corresponding data in (a) and (b),
				in which the blurring of propagation features arising from NNN coupling and
				Floquet driving can be seen more clearly.}
			\label{otocxx}
		\end{figure*}
		
		\begin{figure*}[t]     		
			\includegraphics[width=0.6\textwidth]{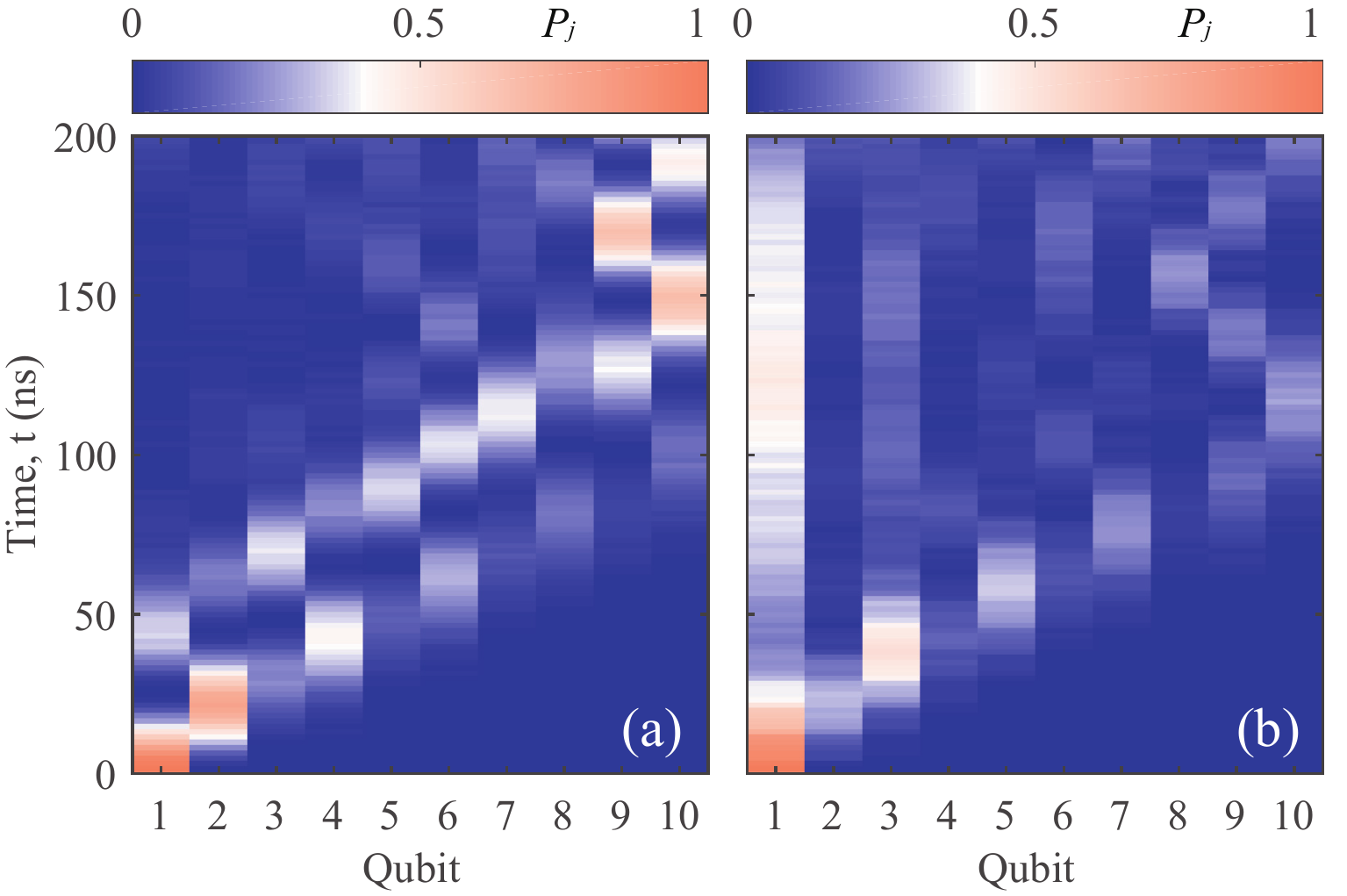}
			\caption{Quench dynamics of the SSH model. A driving amplitude
				$\varepsilon/2\pi=156$~MHz is used, and the photon is initially placed at the
				leftmost qubit $Q_1$. (a) The ac magnetic flux is applied on $Q_3,\ Q_4,\ Q_7$,
				and $Q_8$, where the effective time-independent Hamiltonian is topologically 
				trivial. The photon can almost propagate freely. (b) The ac magnetic flux is
				applied on $Q_2,\ Q_3,\ Q_6$, $Q_7$, and $Q_9$, where the effective
				time-independent Hamiltonian is topologically nontrivial. The photon remains
				localized at the left edge. Each point is the average result of $8,000$
				single-shot readouts.}
			\label{fig_ssh}
		\end{figure*}


\begin{thebibliography}{64}
	
	\makeatletter \providecommand \@ifxundefined [1]{%
		\@ifx{#1\undefined} }%
	\providecommand \@ifnum [1]{%
		\ifnum #1\expandafter \@firstoftwo \else \expandafter \@secondoftwo \fi }%
	\providecommand \@ifx [1]{%
			\ifx #1\expandafter \@firstoftwo \else \expandafter \@secondoftwo \fi }%
		\providecommand \natexlab [1]{#1}%
		\providecommand \enquote  [1]{``#1''}%
		\providecommand \bibnamefont  [1]{#1}%
		\providecommand \bibfnamefont [1]{#1}%
		\providecommand \citenamefont [1]{#1}%
		\providecommand \href@noop [0]{\@secondoftwo}%
		\providecommand \href [0]{\begingroup \@sanitize@url \@href}%
		\providecommand \@href[1]{\@@startlink{#1}\@@href}%
		\providecommand \@@href[1]{\endgroup#1\@@endlink}%
		\providecommand \@sanitize@url [0]{\catcode `\\12\catcode `\$12\catcode `\&12\catcode
			`\#12\catcode `\^12\catcode `\_12\catcode `\%12\relax}%
		\providecommand \@@startlink[1]{}%
		\providecommand \@@endlink[0]{}%
		\providecommand \url  [0]{\begingroup\@sanitize@url \@url }%
		\providecommand \@url [1]{\endgroup\@href {#1}{\urlprefix }}%
		\providecommand \urlprefix  [0]{URL }%
		\providecommand \Eprint [0]{\href }%
		\providecommand \doibase [0]{http://dx.doi.org/}%
		\providecommand \selectlanguage [0]{\@gobble}%
		\providecommand \bibinfo  [0]{\@secondoftwo}%
		\providecommand \bibfield  [0]{\@secondoftwo}%
		\providecommand \translation [1]{[#1]}%
		\providecommand \BibitemOpen [0]{}%
		\providecommand \bibitemStop [0]{}%
		\providecommand \bibitemNoStop [0]{.\EOS\space}%
		\providecommand \EOS [0]{\spacefactor3000\relax}%
		\providecommand \BibitemShut  [1]{\csname bibitem#1\endcsname}%
		\let\auto@bib@innerbib\@empty
		

\bibitem{Shenker2014}S. H. Shenker, and D. Stanford, Black holes and the butterfly
effect, \href{https://link.springer.com/article/10.1007\%2FJHEP03\%282014\%29067}{J.
	High Energy Phys. 03 (2014) 067}.

\bibitem{Roberts2015} D. A. Roberts, D. Stanford, and L. Susskind, Localized shocks,
\href{https://link.springer.com/article/10.1007\%2FJHEP03\%282015\%29051}{J. High Energy
	Phys. 03 (2015) 051}.

\bibitem{Hosur2016} P. Hosur, X. L. Qi, D. A. Roberts, and B. Yoshida, Chaos in quantum
channels \href{https://link.springer.com/article/10.1007/JHEP02(2016)004}{J. High Energy
	Phys. 02 (2016) 004}.

\bibitem{Shen2017} H. Shen, P. Zhang, R. Fan, and H. Zhai, Out-of-time-order correlation at a quantum phase transition,
\href{https://journals.aps.org/prb/abstract/10.1103/PhysRevB.96.054503}{Phys. Rev. B \textbf{96}, 054503 (2017)}. 

\bibitem{Swingle2018}B. Swingle, Unscrambling the physics of out-of-time-order
correlators \href{https://www.nature.com/articles/s41567-018-0295-5}{Nat. Phys. 14, 988
	(2018)}.

\bibitem{Lin2018a}C.-J Lin and O. I. Motrunich, Out-of-time-ordered correlators in a quantum Ising chain,
\href{https://journals.aps.org/prb/abstract/10.1103/PhysRevB.97.144304}{Phys. Rev. B \textbf{97}, 144304 (2018)}.

\bibitem{Lin2018b}C.-J Lin and O. I. Motrunich, Out-of-time-ordered correlators
in short-range and long-range hard-core boson models and in the Luttinger-liquid
model,
\href{https://journals.aps.org/prb/abstract/10.1103/PhysRevB.98.134305}{Phys.
	Rev. B \textbf{98}, 134305 (2018)}.

\bibitem{Joshi2020}M.K. Joshi, A. Elben, B. Vermersch, T. Brydges, C. Maier, P.
Zoller, R. Blatt, and C. F. Roos, Quantum information scrambling in a trapped-ion
quantum simulator with tunable range interactions,
\href{https://journals.aps.org/prl/abstract/10.1103/PhysRevLett.124.240505}{
	Phys. Rev. Lett \textbf{124}, 240505 (2020)}.

\bibitem{Lieb1972}E. H. Lieb and D. W. Robinson, The finite group velocity of quantum
spin systems,
\href{https://link.springer.com/chapter/10.1007/978-3-662-10018-9_25}{Commun. Math.
	Phys. \textbf{28}, 251 (1972)}.

\bibitem{Bravyi2006} S. Bravyi, M. B. Hastings, F. Verstraete, Lieb-Robinson bounds and
the generation of correlations and topological quantum order,
\href{https://journals.aps.org/prl/abstract/10.1103/PhysRevLett.97.050401}{Phys. Rev.
	Lett. \textbf{97}, 050401 (2006)}.

\bibitem{Kitaev2015} A. Kitaev, \textit{A simple model of quantum holography, in
	Proceedings at KITP, 2015} (2015).

\bibitem{Bohrdt2017}A. Bohrdt, C. B. Mendl, M. Endres, and M. Knap, Scrambling and
thermalization in a diffusive quantum many-body system,
\href{https://iopscience.iop.org/article/10.1088/1367-2630/aa719b}{New J. Phys.
	\textbf{19}, 063001 (2017).}

\bibitem{Nahum2018}A. Nahum, S. Vijay, and J. Haah, Operator spreading in random unitary
circuits, \href{https://journals.aps.org/prx/abstract/10.1103/PhysRevX.8.021014}{Phys.
	Rev. X \textbf{8}, 021014 (2018).}

\bibitem{Keyserlingk2018}C. W. von Keyserlingk, T. Rakovszky, F. Pollmann, and S. L.
Sondhi, Operator hydrodynamics, OTOCs, and entanglement growth in systems without
conservation laws,
\href{https://journals.aps.org/prx/abstract/10.1103/PhysRevX.8.021013}{Phys. Rev. X
	\textbf{8}, 021013 (2018).}

%\bibitem{Buluta2009}I. Buluta and F. Nori, Quantum simulators,
%\href{http://science.sciencemag.org/content/326/5949/108}{Science \textbf{326}, 108
%	(2009)}.

%\bibitem{Georgescu2014}I. M. Georgescu, S. Ashhab, and F. Nori, Quantum simulation,
%\href{http://journals.aps.org/rmp/abstract/10.1103/RevModPhys.86.153}{Rev. Mod. Phys.
%	\textbf{86}, 153 (2014)}.

\bibitem{Li2017}J. Li, R. Fan, H. Wang, B. Ye, B. Zeng, H. Zhai, X. Peng, and J. Du,
Measuring out-of-time-order correlators on a nuclear magnetic resonance quantum
simulator, \href{https://journals.aps.org/prx/abstract/10.1103/PhysRevX.7.031011}{Phys.
	Rev. X  \textbf{7}, 031011 (2017)}.

\bibitem{Nie2020}X. Nie, B.-B. Wei, X. Chen, Z. Zhang, X. Zhao, C. Qiu, Y. Tian, Y. Ji,
T. Xin, D. Lu , and J. Li, Experimental observation of equilibrium and dynamical quantum
phase transitions via out-of-time-ordered correlators,
\href{https://journals.aps.org/prl/abstract/10.1103/PhysRevLett.124.250601}{Phys. Rev.
	Lett. \textbf{124}, 250601 (2020)}.

\bibitem{Garttner2017}M. G\"{a}rttner, Justin G. Bohnet, A. Safavi-Naini, M. L.
Wall, J. J. Bollinger, and A. M. Rey, Measuring out-of-time-order correlations
and multiple quantum spectra in a trapped-ion quantum magnet,
\href{https://www.nature.com/articles/nphys4119}{Nat. Phys. \textbf{13}, 781 (2017)}.

\bibitem{Landsman2019} K. A. Landsman, C. Figgatt, T. Schuster, N.M. Linke, B. Yoshida,
N. Y. Yao, and C. Monroe, Verified quantum information scrambling,
\href{https://www.nature.com/articles/s41586-019-0952-6}{Nature (London) \textbf{567},
	61 (2019)}.

\bibitem{Mi2021}X. Mi {\it et al.}, Information scrambling in computationally complex
quantum circuits, \href{https://www.science.org/doi/10.1126/science.abg5029}{Science  \textbf{374}, 1479
	(2021)}.

\bibitem{Braum2021}J. Braum{\" u}ller, A.H. Karamlou, Y. Yanay, B. Kannan, D. Kim, M.
Kjaergaard, A. Melville, B. M. Niedzielski, Y. Sung, A. Veps\"{a}l\"{a}inen, R. Winik,
J. L. Yoder, T. P. Orlando, S. Gustavsson, C. Tahan, and W. D. Oliver, Probing quantum
information propagation with out-of-time-ordered correlators,
\href{https://www.nature.com/articles/s41567-021-01430-w}{Nat. Phys. \textbf{18}, 172
	(2022)}.

\bibitem{rem}Protocols without reversing the time evolution of the system can be
found in Refs.~\cite{Shen2017,Joshi2020}.


\bibitem{Buluta2009}I. Buluta and F. Nori, Quantum simulators,
\href{http://science.sciencemag.org/content/326/5949/108}{Science \textbf{326}, 108
	(2009)}.

\bibitem{Georgescu2014}I. M. Georgescu, S. Ashhab, and F. Nori, Quantum simulation,
\href{http://journals.aps.org/rmp/abstract/10.1103/RevModPhys.86.153}{Rev. Mod. Phys.
	\textbf{86}, 153 (2014)}.


\bibitem{Eckardt2017}A. Eckardt, Colloquium: Atomic quantum gases in periodically driven
optical lattices,
\href{http://journals.aps.org/rmp/abstract/10.1103/RevModPhys.89.011004}{Rev. Mod. Phys.
	\textbf{89}, 011004 (2017)}.

\bibitem{Dunlap1986}D. H. Dunlap, and V. M. Kenkre, Dynamic localization of a charged
particle moving under the influence of an electric field,
\href{https://journals.aps.org/prb/abstract/10.1103/PhysRevB.34.3625}{ Phys. Rev. B
	\textbf{34}, 3625 (1986)}.

\bibitem{Lignier2007}H. Lignier, C. Sias, D. Ciampini, Y. Singh, A. Zenesini, O. Morsch,
and E. Arimondo, Dynamical control of matter-wave tunneling in periodic potentials,
\href{https://journals.aps.org/prl/abstract/10.1103/PhysRevLett.99.220403}{ Phys. Rev.
	Lett \textbf{99}, 220403 (2007)}.

\bibitem{Eckardt2009}A. Eckardt, M. Holthaus, H.Lignier, A. Zenesini, D. Ciampini, O.
Morsch, and E. Arimondo, Exploring dynamic localization with a Bose-Einstein condensate,
\href{https://journals.aps.org/pra/abstract/10.1103/PhysRevA.79.013611}{ Phys. Rev. Lett
	\textbf{79}, 013611 (2009)}.

\bibitem{Dalibard2011}J. Dalibard, F. Gerbier, G. Juzeli\={u}nas, and P. \"{O}hberg,
Colloquium: Artificial gauge potentials for neutral atoms,
\href{http://journals.aps.org/rmp/abstract/10.1103/RevModPhys.83.1523}{Rev. Mod. Phys.
	\textbf{85}, 1523 (2011)}.

\bibitem{Jotzu2014}G. Jotzu, M. Messer, R. Desbuquois, M. Lebrat, T. Uehlinger, D.
Greif, and T. Esslinger, Experimental realization of the topological Haldane model with
ultracold fermions, \href{https://www.nature.com/articles/nature13915}{ Nature
	\textbf{515}, 237 (2014)}.

\bibitem{Wu2018}Y. Wu, L. Yang, M. Gong, Y. Zheng, H. Deng, Z. Yan, Y. Zhao, K. Huang,
A. D. Castellano, W. J. Munro, K. Nemoto, D. Zheng, C. P. Sun, Y. X. Liu, X. Zhu, and L.
Lu, An efficient and compact switch for quantum circuits,
\href{https://www.nature.com/articles/s41534-018-0099-6}{npj Quantum Inf. \textbf{4}, 50
	(2018)}.

\bibitem{Reagor2018}M. Reagor, C. B. Osborn, N. Tezak, A. Staley, G. Prawiroatmodjo, M.
Scheer \textit{et al.}, Demonstration of universal parametric entangling gates on a
multi-qubit lattice, \href{https://advances.sciencemag.org/content/4/2/eaao3603}{Sci.
	Adv. \textbf{4}, eaao3603 (2018)}.

\bibitem{Li2018}X. Li, Y. Ma, J. Han, T. Chen, Y. Xu,W. Cai, H.Wang, Y. P. Song, Z.-Y.
Xue, Z.-Q. Yin, and L. Sun, Perfect quantum state transfer in a superconducting qubit
chain with parametrically tunable couplings,
\href{https://journals.aps.org/prapplied/abstract/10.1103/PhysRevApplied.10.054009} 
{Phys. Rev. Applied \textbf{10}, 054009 (2018)}.

\bibitem{Cai2019}W. Cai, J. Han, F. Mei, Y. Xu, Y. Ma, X. Li, H. Wang, Y. P. Song, Z.-Y.
Xue, Z. Yin, S. Jia, and L. Sun, Observation of topological magnon insulator states in a
superconducting circuit,
\href{https://link.aps.org/doi/10.1103/PhysRevLett.123.080501}{Phys. Rev. Lett.
	\textbf{124}, 080501 (2019)}.

\bibitem{Salathe2015}Y. Salath\'{e}, M. Mondal, M. Oppliger, J. Heinsoo, P. Kurpiers, A.
Poto\v{c}nik, A. Mezzacapo, U. Las Heras, L. Lamata, E. Solano, S. Filipp, and A.
Wallraff, Digital quantum simulation of spin models with circuit quantum
electrodynamics,
\href{https://journals.aps.org/prx/abstract/10.1103/PhysRevX.5.021027}{Phys. Rev. X 
	\textbf{5}, 021027 (2015)}.

\bibitem{Barends2015}R. Barends, L. Lamata, J. Kelly, L. Garc\'{\i}a-\'{A}lvarez, A. G.
Fowler, A Megrant, E Jeffrey, T. C. White, D. Sank, J. Y. Mutus, B. Campbell, Yu Chen,
Z. Chen, B. Chiaro, A. Dunsworth, I.-C. Hoi, C. Neill, P. J. J. O'Malley, C. Quintana,
P. Roushan, A. Vainsencher, J. Wenner, E. Solano, and John M. Martinis, Digital quantum
simulation of fermionic models with a superconducting circuit,
\href{https://www.nature.com/articles/ncomms8654}{Nat. Commun.  \textbf{6}, 7654
	(2015)}.

\bibitem{Zhong2016}Y. P. Zhong,  D. Xu, P. Wang, C. Song, Q. J. Guo, W. X. Liu, K. Xu,
B. X. Xia, C.-Y. Lu, S. Han, J.-W. Pan, and H. Wang, Emulating anyonic fractional
statistical behavior in a superconducting quantum circuit,
\href{https://journals.aps.org/prl/abstract/10.1103/PhysRevLett.117.110501}{Phys. Rev.
	Lett.  \textbf{117}, 110501 (2016)}.

\bibitem{Flurin2017}E. Flurin, V. V. Ramasesh, S. Hacohen-Gourgy, L. S. Martin, N. Y.
Yao, and I. Siddiqi, Observing topological invariants using quantum walks in
superconducting circuits,
\href{https://journals.aps.org/prx/abstract/10.1103/PhysRevX.7.031023}{ Phys. Rev. X
	\textbf{7}, 031023 (2017)}.

\bibitem{Roushan2017}P. Roushan, C. Neill, J. Tangpanitanon, V. M. Bastidas, A. Megrant,
R. Barends, Y. Chen, Z. Chen, B. Chiaro, A. Dunsworth, A. Fowler, B. Foxen, M. Giustina,
E. Jeffrey, J. Kelly, E. Lucero, J. Mutus, M. Neeley, C. Quintana, D. Sank, A.
Vainsencher, J. Wenner, T. White, H. Neven, D. G. Angelakis, and J. Martinis,
Spectroscopic signatures of localization with interacting photons in superconducting
qubits, \href{http://science.sciencemag.org/content/358/6367/1175}{Science \textbf{358},
	1175 (2017)}.

\bibitem{Xu2018}K. Xu, J. J. Chen, Y. Zeng, Y. R. Zhang, C. Song, W. X. Liu, Q. J. Guo,
P. F. Zhang, D. Xu, H. Deng, K. Q. Huang, H. Wang, X. B. Zhu, D. N. Zheng, and H. Fan,
Emulating many-body localization with a superconducting quantum processor,
\href{https://journals.aps.org/prl/abstract/10.1103/PhysRevLett.120.050507}{Phys. Rev.
	Lett.  \textbf{120}, 050507 (2018)}.

\bibitem{Song2018}C. Song, D. Xu, P. Zhang, J. Wang, Q. Guo, W. Liu, K. Xu, H. Deng, K.
Huang, D. Zheng, S.-B. Zheng, H. Wang, X. Zhu, C.-Y. Lu, and J.-W. Pan, Demonstration of
topological robustness of anyonic braiding statistics with a superconducting quantum
circuit,
\href{https://journals.aps.org/prl/abstract/10.1103/PhysRevLett.121.030502}{Phys. Rev.
	Lett.  \textbf{121}, 030502 (2018)}.

\bibitem{Yan2019}Z. Yan, Y. R. Zhang, M. Gong, Y. Wu, Y. Zheng, S. Li, C. Wang, F.
Liang, J. Lin, Y. Xu, C. Guo, L. Sun, C. Z. Peng, K. Xia, H. Deng, H. Rong, J. Q. You,
F. Nori, H. Fan, X. Zhu, and J.-W. Pan, Strongly correlated quantum walks with a
12-qubit superconducting processor,
\href{https://science.sciencemag.org/content/364/6442/753}{Science  \textbf{364}, 753
	(2019)}.

\bibitem{Ma2019}R. Ma, B. Saxberg, C. Owens, N. Leung, Y. Lu, J. Simon, and D. I.
Schuster, A dissipatively stabilized Mott insulator of photons,
\href{https://www.nature.com/articles/s41586-019-0897-9}{Nature  \textbf{566}, 51
	(2019)}.

\bibitem{Ye2019}Y. Ye, Z.-Y. Ge, Y. Wu, S. Wang, M. Gong, Y.-R. Zhang, Q. Zhu, R. Yang,
S. Li, F. Liang, J. Lin, Y. Xu, C. Guo, L. Sun, C. Cheng, N. Ma, Z. Y. Meng, H. Deng, H.
Rong, C.-Y. Lu, C.-Z. Peng, H. Fan, X. Zhu, and J.-W. Pan, Propagation and localization
of collective excitations on a 24-qubit superconducting processor,
\href{https://link.aps.org/doi/10.1103/PhysRevLett.123.050502}{Phys. Rev. Lett.
	\textbf{123}, 050502 (2019)}.

\bibitem{Guo2019}X.-Y Guo, C. Yang, Y. Zeng, Y. Peng, H.-K Li, H. Deng, Y.-R Jin, S.
Chen, D.-N Zheng, and H. Fan, Observation of a dynamical quantum phase transition by a
superconducting qubit simulation,
\href{https://link.aps.org/doi/10.1103/PhysRevApplied.11.044080}{Phys. Rev. Applied
	\textbf{11}, 044080 (2019)}.

\bibitem{Arute2019}F. Arute, K. Arya, R. Babbush, D. Bacon, J. C. Bardin, R. Barends, R.
Biswas, S. Boixo, F. G.Brandao, D. A. Buell \textit{et al}., Quantum supremacy using a
programmable superconducting processor,
\href{https://doi.org/10.1038/s41586-019-1666-5}{Nature (London) \textbf{574}, 505
	(2019)}.

\bibitem{Xu2020}K. Xu, Z.-H Sun, W. Liu, Y.-R Zhang, H. Li, H. Dong, W. Ren, P. Zhang,
F. Nori, D. Zheng, H. Fan, H. Wang, Probing dynamical phase transitions with a
superconducting quantum simulator,
\href{https://advances.sciencemag.org/content/6/25/eaba4935.full}{Sci. Adv. \textbf{6},
	eaba4935 (2020)}.

\bibitem{Guo2021}X.-Y Guo, Z.-Y. Ge, H. Li, Z. Wang, Y.-R. Zhang, P. Song, Z.  Xiang, X.
Song, Y. Jin, L. Lu, K. Xu, D. Zheng, and H. Fan, Observation of Bloch oscillations and
Wannier-Stark localization on a superconducting quantum processor,
\href{https://www.nature.com/articles/s41534-021-00385-3}{npj Quantum Inf. \textbf{7},
	51 (2021)}.

\bibitem{Guoqj2021}Q. Guo, C. Cheng, Z.-H. Sun, Z. Song, H. Li, Z. Wang, W. Ren, H.
Dong, D. Zheng, Y.-R. Zhang, R. Mondaini, H. Fan, and H. Wang, Observation of
energy-resolved many-body localization,
\href{https://www.nature.com/articles/s41567-020-1035-1}{Nat. Phys. \textbf{17}, 234
	(2021)}.

\bibitem{SM} See Supplemental Material.

\bibitem{Gong2021}M. Gong, S. Wang, C. Zha, M.-C. Chen, H.-L. Huang, Y. Wu, Q. Zhu, Y.
Zhao, S. Li, S. Guo, H. Qian, Y. Ye, F. Chen, C. Ying, J. Yu, D. Fan, D. Wu, H. Su, H.
Deng, H. Rong, K. Zhang, S. Cao, J. Lin, Y. Xu, L. Sun, C. Guo, N. Li, F. Liang, V. M.
Bastidas, K. Nemoto, W. J. Munro, Y.-H. Huo, C.-Y. Lu, C.-Z. Peng, X. Zhu, and J.-W.
Pan, Quantum walks on a programmable two-dimensional 62-qubit superconducting processor,
\href{https://science.sciencemag.org/content/372/6545/948}{Science \textbf{373}, 948
	(2021)}.

\bibitem{Underwood2012}M. S. Underwood, and D. L. Feder, Bose-Hubbard model for
universal quantum-walk-based computation,
\href{https://link.aps.org/doi/10.1103/PhysRevA.85.052314}{Phys. Rev. A \textbf{85},
	052314 (2012)}.

\bibitem{Childs2013}A. M. Childs, D. Gosset, and Z. Webb, Universal computation by
multiparticle quantum walk,
\href{https://science.sciencemag.org/content/339/6121/791}{Science \textbf{339}, 791
	(2013)}.

\bibitem{McKay2017}D. C. McKay, C. J. Wood, S. Sheldon, J. M. Chow, and J. M. Gambetta,
Efficient Z gates for quantum computing,
\href{https://link.aps.org/doi/10.1103/PhysRevA.96.022330}{Phys. Rev. A \textbf{96},
	022330 (2017)}.

\bibitem{Polkovnikov2011}A. Polkovnikov, K. Sengupta, A. Silva, and M. Vengalattore,
Colloquium: Nonequilibrium dynamics of closed interacting quantum systems,
\href{https://link.aps.org/doi/10.1103/RevModPhys.83.863}{Rev. Mod. Phys. \textbf{83},
	863 (2011)}.

\bibitem{Marcuzzi2013}M. Marcuzzi, J. Marino, A. Gambassi, and A. Silva,
Prethermalization in a nonintegrable quantum spin chain after a quench,
\href{https://link.aps.org/doi/10.1103/PhysRevLett.111.197203}{Phys. Rev. Lett.
	\textbf{111}, 197203 (2013)}.

\bibitem{Bansil2016}A. Bansil, H. Lin, and T. Das, Colloquium: Topological band theory,
\href{http://journals.aps.org/rmp/abstract/10.1103/RevModPhys.88.021004}{Rev. Mod. Phys.
	\textbf{88}, 021004 (2016)}.

\bibitem{Su1979}W. P. Su, J. R. Schrieffer, and A. J. Heeger, Solitons in Polyacetylene,
\href{https://link.aps.org/doi/10.1103/PhysRevLett.42.1698}{Phys. Rev. Lett.
	\textbf{42}, 1698 (1979)}.

\end{thebibliography}

\begin{thebibliography}{64}
	
	\makeatletter \providecommand \@ifxundefined [1]{%
		\@ifx{#1\undefined} }%
	\providecommand \@ifnum [1]{%
		\ifnum #1\expandafter \@firstoftwo \else \expandafter \@secondoftwo \fi }%
	\providecommand \@ifx [1]{%
		\ifx #1\expandafter \@firstoftwo \else \expandafter \@secondoftwo \fi }%
	\providecommand \natexlab [1]{#1}%
	\providecommand \enquote  [1]{``#1''}%
	\providecommand \bibnamefont  [1]{#1}%
	\providecommand \bibfnamefont [1]{#1}%
	\providecommand \citenamefont [1]{#1}%
	\providecommand \href@noop [0]{\@secondoftwo}%
	\providecommand \href [0]{\begingroup \@sanitize@url \@href}%
	\providecommand \@href[1]{\@@startlink{#1}\@@href}%
	\providecommand \@@href[1]{\endgroup#1\@@endlink}%
	\providecommand \@sanitize@url [0]{\catcode `\\12\catcode `\$12\catcode `\&12\catcode
		`\#12\catcode `\^12\catcode `\_12\catcode `\%12\relax}%
	\providecommand \@@startlink[1]{}%
	\providecommand \@@endlink[0]{}%
	\providecommand \url  [0]{\begingroup\@sanitize@url \@url }%
	\providecommand \@url [1]{\endgroup\@href {#1}{\urlprefix }}%
	\providecommand \urlprefix  [0]{URL }%
	\providecommand \Eprint [0]{\href }%
	\providecommand \doibase [0]{http://dx.doi.org/}%
	\providecommand \selectlanguage [0]{\@gobble}%
	\providecommand \bibinfo  [0]{\@secondoftwo}%
	\providecommand \bibfield  [0]{\@secondoftwo}%
	\providecommand \translation [1]{[#1]}%
	\providecommand \BibitemOpen [0]{}%
	\providecommand \bibitemStop [0]{}%
	\providecommand \bibitemNoStop [0]{.\EOS\space}%
	\providecommand \EOS [0]{\spacefactor3000\relax}%
	\providecommand \BibitemShut  [1]{\csname bibitem#1\endcsname}%
	\let\auto@bib@innerbib\@empty
	
	
	\bibitem{Koch2007}J. Koch, T. M. Yu, J. Gambetta, A. A. Houck, D. I. Schuster, J. Majer, A. Blais,
	M. H. Devoret, S. M. Girvin, and R. J. Schoelkopf, “Charge-insensitive qubit
	design derived from the cooper pair box,” Phys. Rev. A \textbf{76}, 042319 (2007).
	
	\bibitem{Barends2013}R. Barends, J. Kelly, A. Megrant, D. Sank, E. Jeffrey, Y. Chen, Y. Yin, B.
	Chiaro, J. Mutus, C. Neill, P. O’Malley, P. Roushan, J. Wenner, T. C. White,
	A. N. Cleland, and J. M. Martinis, Coherent Josephson qubit suitable for
	scalable quantum integrated circuits, Phys. Rev. Lett. \textbf{111}, 080502 (2013).
	
	\bibitem{Chen2014}Z. Chen, A. Megrant, J. Kelly, R. Barends, J. Bochmann,
	Y. Chen, B. Chiaro, A. Dunsworth, E. Jeffrey, J. Y.
	Mutus, P. J. J. O‘Malley, C. Neill, P. Roushan, D. Sank,
	A. Vainsencher, J. Wenner, T. C. White, A. N. Cleland, and J. M. Martinis,
	Fabrication and characterization of aluminum airbridges for superconducting microwave circuits,
	\href{https://aip.scitation.org/doi/10.1063/1.4863745}{Appl. Phys. Lett. \textbf{104}, 052602 (2014)}.
	
	\bibitem{Guo2021}X.-Y Guo, Z.-Y. Ge, H. Li, Z. Wang, Y.-R. Zhang, P. Song, Z.  Xiang, X. Song, Y. Jin, L. Lu, K. Xu, D. Zheng, and H. Fan,
	Observation of Bloch oscillations and Wannier-Stark localization on a superconducting quantum processor,
	\href{https://www.nature.com/articles/s41534-021-00385-3}{npj Quantum Inf. \textbf{7}, 51 (2021)}.
	
	\bibitem{Yan2019}Z. Yan, Y. R. Zhang, M. Gong, Y. Wu, Y. Zheng, S. Li, C. Wang,
	F. Liang, J. Lin, Y. Xu, C. Guo, L. Sun, C. Z. Peng, K. Xia, H. Deng, H. Rong, J. Q. You, F. Nori, H. Fan, X. Zhu, and J.-W. Pan,
	Strongly correlated quantum walks with a 12-qubit superconducting processor,
	\href{https://science.sciencemag.org/content/364/6442/753}{Science  \textbf{364}, 753 (2019)}.
	
	\bibitem{Reed2013}M. D. Reed, {\it Entanglement and Quantum Error Correction with
		Superconducting Qubits}, Ph.D. thesis (2013).
	
	\bibitem{Ku2020}Jaseung Ku, Xuexin Xu, Markus Brink, David C. McKay, Jared B.
	Hertzberg, Mohammad H. Ansari, and B. L. T. Plourde. Suppression of Unwanted ZZ 
	Interactions in a Hybrid Two-Qubit System, Phys. Rev. Lett. \textbf{125}, 200504 (2020).
	
	\bibitem{Song2017}Chao Song, Kai Xu, Wuxin Liu, Chui-ping Yang, Shi-Biao Zheng,,
	Hui Deng, Qiwei Xie, Keqiang Huang, Qiujiang Guo, Libo Zhang, Pengfei Zhang, Da
	Xu, Dongning Zheng, Xiaobo Zhu, H. Wang, Y.-A. Chen, C.-Y. Lu, Siyuan Han, and
	Jian-Wei Pan, 10-Qubit Entanglement and Parallel Logic Operations with a
	Superconducting Circuit, Phys. Rev. Lett. \textbf{119}, 180511 (2017).
	
	\bibitem{Ye2019}Y. Ye, Z.-Y. Ge, Y. Wu, S. Wang, M. Gong, Y.-R. Zhang,
	Q. Zhu, R. Yang, S. Li, F. Liang, J. Lin, Y. Xu, C. Guo,
	L. Sun, C. Cheng, N. Ma, Z. Y. Meng, H. Deng, H. Rong,
	C.-Y. Lu, C.-Z. Peng, H. Fan, X. Zhu, and J.-W. Pan,
	Propagation and localization of collective excitations on a 24-qubit superconducting processor,
	\href{https://link.aps.org/doi/10.1103/PhysRevLett.123.050502}{Phys. Rev. Lett. \textbf{123}, 050502 (2019)}.
	
	\bibitem{McKay2017}D. C. McKay, C. J. Wood, S. Sheldon, J. M. Chow, and J. M. Gambetta,
	Efficient Z gates for quantum computing,
	\href{https://link.aps.org/doi/10.1103/PhysRevA.96.022330}{Phys. Rev. A \textbf{96}, 022330 (2017)}.
	
	\bibitem{Braum2021}J. Braum{\" u}ller, A.H. Karamlou, Y. Yanay, B. Kannan, D. Kim, M.
	Kjaergaard, A. Melville, B. M. Niedzielski, Y. Sung, A. Veps\"{a}l\"{a}inen, R. Winik,
	J. L. Yoder, T. P. Orlando, S. Gustavsson, C. Tahan, and W. D. Oliver, Probing quantum
	information propagation with out-of-time-ordered correlators,
	\href{https://arxiv.org/abs/2102.11751}{arXiv:2102.11751}.
	
	\bibitem{Su1979}W. P. Su, J. R. Schrieffer, and A. J. Heeger,
	Solitons in Polyacetylene,
	\href{https://link.aps.org/doi/10.1103/PhysRevLett.42.1698}{Phys. Rev. Lett. \textbf{42}, 1698 (1979)}.
	
\end{thebibliography}
\end{document}